\documentclass[final,3p,twocolumn,times]{elsarticle}
\usepackage{comment}
\usepackage{subfig}
\usepackage{url}
\usepackage{hyperref}
\usepackage{graphics}
\usepackage{graphicx}
\usepackage{amsmath}
\usepackage{bm}
\usepackage{booktabs}
\usepackage{balance}
\usepackage{float}
\usepackage{amsfonts}
\usepackage[dvipsnames]{xcolor}
\usepackage[export]{adjustbox}
\usepackage[ruled,  linesnumbered, commentsnumbered, longend]{algorithm2e}

\definecolor{tablered}{rgb}{0.8, 0.25, 0.33}
\definecolor{tablegreen}{rgb}{0.0, 0.26, 0.15}

\begin{document}

\journal{Computer Communications}

\begin{frontmatter}
\title{Edge Computing vs Centralized Cloud: Impact of Communication Latency on the Energy Consumption of LTE Terminal Nodes}

\author[unifi,dip]{Chiara~Caiazza}
\ead{chiara.caiazza@unifi.it}
\author[sup]{Silvia~Giordano}
\ead{silvia.giordano@supsi.ch}
\author[iit]{Valerio~Luconi}
\ead{valerio.luconi@iit.cnr.it}
\author[dip]{Alessio~Vecchio\corref{cor1}}
\ead{alessio.vecchio@unipi.it}

\cortext[cor1]{Corresponding author}
\address[dip]{Dip. di Ing. dell'Informazione, Universit\`a di Pisa, Largo L. Lazzarino 1, 56122 Pisa, Italy}
\address[unifi]{University of Florence, Italy}
\address[sup]{Dip. di Tecnologie Innovative, SUPSI, Polo Universitario Lugano -- Campus Est, Via la Santa 1, CH-6962 Lugano-Viganello, Switzerland}
\address[iit]{Istituto di Informatica e Telematica, Consiglio Nazionale delle Ricerche, Via G. Moruzzi, 1, 56124 Pisa, Italy}

\date{August 2021}

\begin{abstract}

Edge computing brings several advantages, such as reduced latency, increased bandwidth, and improved locality of traffic. One aspect that is not sufficiently understood is to what extent the different communication latency experienced in the edge-cloud continuum impacts on the energy consumption of clients. We studied the energy consumption of a request-response communication scheme when an LTE node communicates with edge-based or cloud-based servers. Results show that the reduced latency of edge servers bring significant benefits in terms of energy consumption. Experiments also show how the energy savings brought by edge computing are influenced by the prevalent direction of data transfer (upload vs download), load of the server, and daytime/nighttime operation.

\end{abstract}

\end{frontmatter}

\section{Introduction}

With edge computing, processing and storage capabilities are moved from centralized data centers to the periphery of the network, so that they can be close to the end users. The aim of edge computing is to provide low latency and high bandwidth that can be exploited to improve existing applications or to build new ones. The migration of cloud services to the edge can bring benefits in several domains. The reduced latency improves the performance of applications such as online gaming~\cite{8685768}, augmented reality~\cite{7980118}, and connected vehicles~\cite{8515152}. Real-time video analytics~\cite{8567664} and other traffic-intensive applications can benefit from higher bandwidth. Additionally, the edge computing paradigm allows keeping the traffic local to the access network without the need to transit through the public Internet. This may reduce the load on the core network, especially when bandwidth-intensive applications are considered. In this paper, we use the \emph{edge computing} term according to its broadest meaning: an architectural solution where computing and storage resources are in proximity of the end users, without necessarily assuming that such resources must be co-located with specific elements of the network architecture, such as eNodeB or RAN infrastructure (\cite{giust2018mec} describes some deployment options). The rationale is that we are interested in evaluating edge computing as an architectural design solution, without specific implementation and/or deployment details.

Both edge and cloud solutions can be beneficial for applications running on smart things~\cite{8969038}. 
Many smart devices collect data from the environment, examples include the air quality sensors in a smart city \cite{7740617}, intelligent cameras \cite{MASUD2020215}, and e-health monitoring appliances~\cite{yang2016iot}. Produced information is transmitted to the server side, where it is aggregated, analyzed, and stored. The server side is typically implemented as a set of cloud services, to cope with fluctuations in computing and storage demand~\cite{RAY201635,7116451,botta2016integration}. 
In the last years, also the adoption of the edge computing in combination with smart devices received significant attention~\cite{samie2016computation,morabito2018consolidate,7469991}. Pushing computing, storage, and intelligence closer to the sources of information demonstrated its benefits in several domains, such as manufacturing~\cite{8466364}, wearable cognitive assistance~\cite{10.1145/2967360.2967369}, and smart transportation~\cite{10.1145/3132211.3134446}. Investigation focused on several aspects of the edge vs cloud relationship, such as service deployment, image and stream processing, and task placement~\cite{Savaglio19:iot, Heredia19:edge, Silva19:investigating, Nikolaou19:evaluation, Luckow21:exploring}.

The advantages of edge computing have almost always been quantified in terms of improved response times. An aspect that did not receive enough attention is the impact of edge computing on the energy consumption of client devices, which is definitely important as such devices are typically battery-operated. In this paper, we answer the following research question: how much energy is saved on client devices when using edge computing compared to cloud computing during communication? The study focuses on a request-response protocol, as this is the way of operating of the majority of applications that interact with the cloud (and hence also with the edge).

The client, which issues the requests, is supposed to be energy-constrained and wirelessly connected, as it happens for many smart and/or mobile devices. The server is supposed to be placed either in the edge or in the cloud, with significant differences in terms of communication latency. The communication between the two application components (client and server) is assumed to be based on LTE, as this is the technology currently used for a significant fraction of machine-to-machine communication.
To answer the above research question, this work brings the following technical contributions:
\begin{itemize}

    \item An analytical model that describes the energy consumption of a request-response protocol is provided. The model, differently from existing literature, allows application designers to understand how the various phases of the protocol, such as transmitting the data or waiting for a reply, affect the overall energy consumption when communicating with an edge server or a cloud one. 
    
    \item The energy needed on client devices, when communicating with edge- and cloud-based servers, is studied when varying the most important parameters of operations. Results show that edge computing, most of the time, can reduce the energy consumption of client devices. Our results concern a request-response communication scheme, whereas existing literature on the topic mostly covers offloading of computation. 

    \item The analytical model is fed with network traces collected in the real world to incorporate in the evaluation the complexities of the TCP protocol. Results show that edge computing can bring a significant reduction of the energy consumed on the client side, up to 55\%, when the request-response communication scheme operates on top of a connection-oriented protocol. Experiments also show how the energy savings brought by edge computing can be affected by the prevalent direction of data transfer (upload vs download), load of the server, and daytime/nighttime operation. 
    
\end{itemize}

The rest of this paper is organized as follows.
The scenario of operation and the communication scheme are presented in Section~\ref{sec:scenario}, while the model of the LTE interface is given in Section~\ref{sec:power}. The energy consumption of the considered communication scheme is computed in Section~\ref{sec:consumption}, while results for connectionless and connection-oriented communication are shown in Sections~\ref{sec:udpresults} and~\ref{sec:tcpresults}, respectively. Section~\ref{sec:discussion} discusses the main findings and the limitations of this study. Section~\ref{sec:relwork} summarizes the related work. Finally, Section~\ref{sec:conclusions} concludes this work.
To conclude, Table \ref{table:symbols} summarises the symbols used in the rest of this paper.

\begin{table*}[t]
    \centering
    \caption{The list of symbols used.}
    \begin{tabular}{lp{0.9\textwidth}}
        \toprule
    
        $T_{I}$    & The application period (the interval between a request and the next one)                   \\
        $T_{TX}$   & The time needed to transmit the data to the server                                         \\ 
        $T_{RX}$   & The time needed to receive the data from the server                                        \\
        $T_{ELAB}$ & The elaboration time of the server                                                                      \\
        $T_W$      & The time between the end of transmission and the beginning of reception                   \\
        $T_Q$      & The time elapsed between the end of $T_{RX}$ and the end of the current application period \\
        $E_{I}$    & The energy consumption during $T_I$, i.e. during a single request-response cycle                                                        \\
        $E_{TX}$   & The energy consumption during $T_{TX}$                                                     \\
        $E_{RX}$   & The energy consumption during $T_{RX}$                                                     \\
        $E_W$      & The energy consumption during $T_W$                                                        \\
        $E_Q$      & The energy consumption during $T_Q$                                                        \\
        $T_C$      & The time spent in CR before moving to SHORT DRX in the absence of packets sent or received                        \\
        $T_S$      & The time spent in SHORT DRX before moving to LONG DRX in the absence of packets sent or received                            \\
        $T_L$      & The time spent in LONG DRX before moving to IDLE in the absence of packets sent or received                             \\
        $T_{PROM}$ & The time needed to promote the interface from IDLE to CR                                   \\
        $P_C$      & The power consumption during $T_C$                                                         \\
        $P_S$      & The power consumption during $T_S$                                                         \\
        $P_L$      & The power consumption during $T_L$                                                         \\
        $P_{PROM}$ & The power consumption during $T_{PROM}$                                                    \\
        
        \bottomrule
    \end{tabular}
    \label{table:symbols}
\end{table*}

\section{Scenario of Operation and Method}
\label{sec:scenario}

\begin{figure}[!t]
    \centering
    \includegraphics[width=0.45\textwidth]{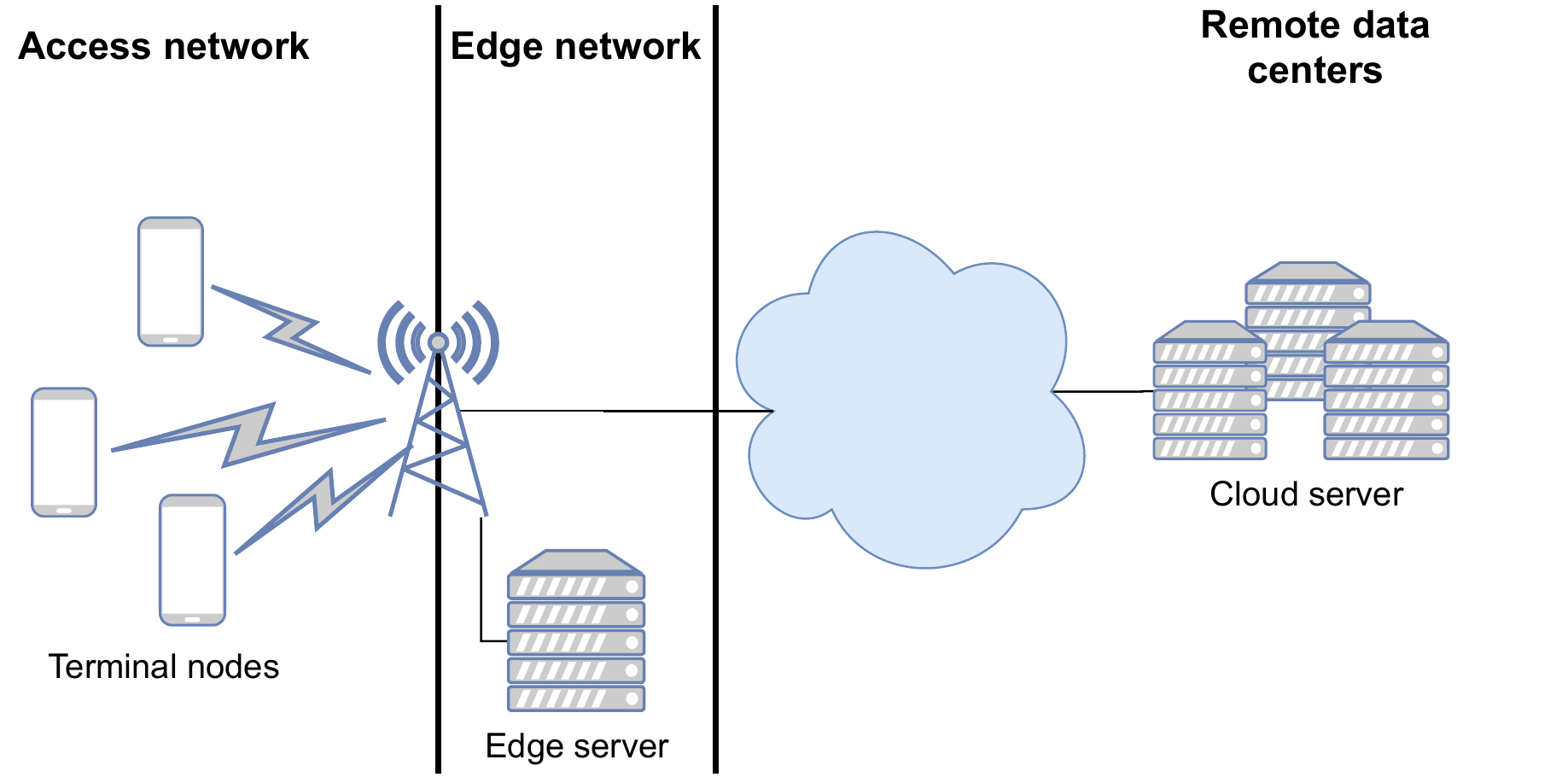}
    \caption{A modern network architecture including an access network, an edge network, and remote data centers.}
    \label{fig:MECArchitecture}
\end{figure}

We assume a scenario of operation like the one depicted in Figure~\ref{fig:MECArchitecture}. The scenario comprises a typical network architecture where an access network provides connectivity to the terminal nodes, and an edge network hosts the edge servers and connects them to the base stations (BSs). The edge network is connected, directly or indirectly, to the public Internet from which remote cloud servers can be reached. 
Terminal nodes communicate with application components that can be executed either on edge servers or cloud servers. Depending on the placement of the application, various advantages can be achieved. When applications are running on edge servers, communication is generally characterized by reduced latency and higher bandwidth. On the contrary, the use of a remote cloud brings benefits in terms of available computing power and ease of operation. For simplicity, we consider a scenario with just one edge server and one cloud server. Since our aim is to study the energy consumption solely on the client side, and how this energy consumption depends on the latency, a scenario including multiple edge or cloud servers would add unnecessary complexity to the problem, as, from our perspective, different edge or cloud servers are only characterized by different latency towards the client.

 \begin{figure}[!t]
    \centering
    \includegraphics[width=0.45\textwidth]{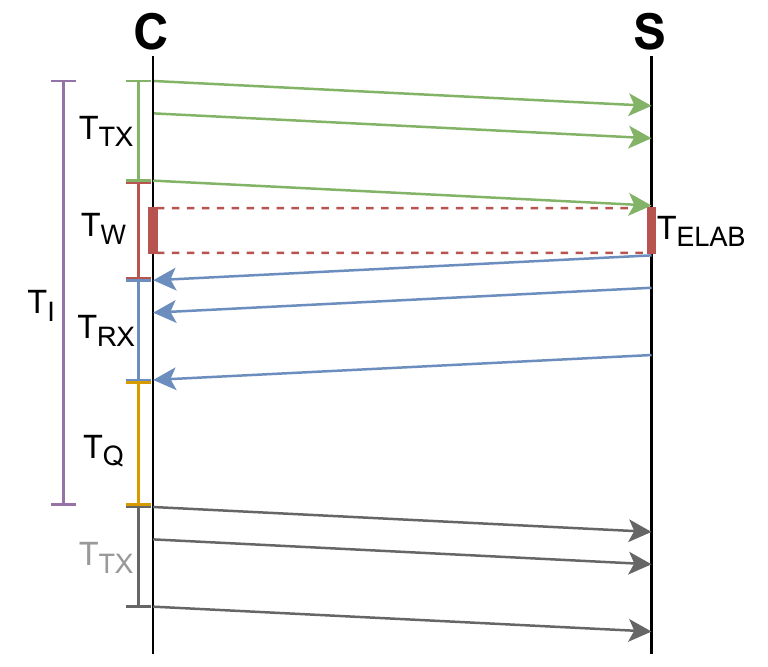}
    \caption{The interaction between the client (C) and the server (S) during the period $T_I$.}
    \label{fig:Application}
\end{figure}

In this scenario, we consider a client-server application, represented in Figure~\ref{fig:Application}. The client (C) runs on terminal nodes, while the server (S) can run either on edge or on cloud machines. The application operates according to a request-response model, whose period is equal to $T_I$. Such periodic behavior can be found in many classes of applications, e.g., a smart camera that periodically sends the collected data to a central repository or a node that checks for incoming messages or updated information according to a polling-based approach. A Machine Type Communication (MTC) traffic pattern that is of periodic nature is indicated in \cite{6629847} as one of the two recurrent schemes. The other one is the event-based communication pattern, which is however outside the scope of this paper. At the beginning of each period, C sends a request to the server. This operation requires a time $T_{TX}$ that depends on multiple factors: the amount of data to be transferred, the bandwidth of the wireless segment, and the communication latency. We assume that S is always idle, thus it is always waiting for incoming requests from C, and that the received request is immediately processed in a fixed amount of time $T_{ELAB}$ (in Section~\ref{sec:multiple-clients}, the impact of an increasing load on the server will be considered). S then sends back the response to C, e.g., a confirmation message that the request has been correctly processed. From the client's perspective, $T_W$ is the interval between the end of the transmission of the last byte of the request and the start of the reception of the first byte of the response. The reception of the response has a duration of $T_{RX}$. Finally, C remains idle for a time $T_Q$ before starting a new request-response cycle.

\subsection{Method}

To estimate the energy consumption of the request-response communication scheme, a power model of the LTE interface of client devices is needed. The power model defines the operational states of the interface, the possible transitions between such states, and the power needed in every state. The transition between the operational states is driven by transmissions and reception of packets, or their lack of.

The model is used to study the energy consumption when communication takes place on top of a connection-less protocol or on top of a connection-oriented one. In the first case, the study is purely analytical as the communication pattern is simple and an analytical approach allows us to understand the impact of network conditions (namely the delay), and operational parameters on the different phases of the communication pattern. In the second case, the complexities of transport-level protocols (such as TCP) do not allow a purely analytical approach. Thus, the energy consumption is estimated by feeding the power model of the interface with data extracted from real world experiments.

\section{Power model of an LTE interface}
\label{sec:power}

\begin{figure}[!t]
    \centering
    \includegraphics[width=0.45\textwidth]{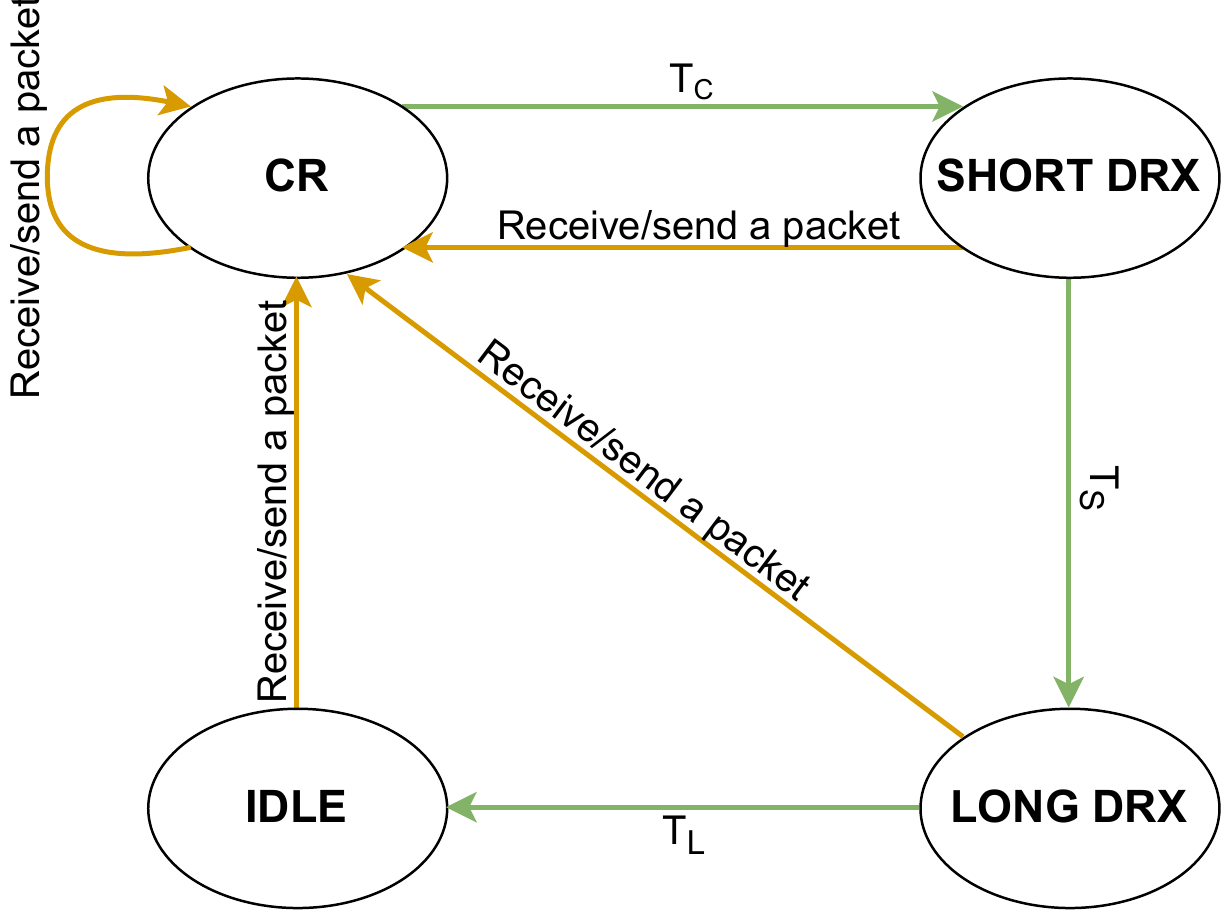}
    \caption{The finite state machine of an LTE module.}
    \label{fig:LTEStateMachine}
\end{figure}

The power model of the LTE interface we consider is the one described in \cite{10.1145/2796314.2745875}, both in terms of required power and duration of transitions (the technique was originally proposed in \cite{10.1145/2307636.2307658}). The interface operates as a Finite State Machine with four states: CR (Continuous Reception), SHORT DRX (Short Discontinuous Reception), LONG DRX (Long Discontinuous Reception), and IDLE. The possible transitions between states are shown in Figure~\ref{fig:LTEStateMachine}. Overall, the LTE interface stays in the more energy-hungry state CR when it has to send and receive packets, and moves to the most energy efficient state IDLE when there is no traffic. DRX states are intermediate states used for transitioning in less energy-hungry states gradually, as specified in the following.

In detail, every time the interface has to send or receive a packet, it enters the CR state, which is the state with the highest power consumption. Once the transmission is completed, the LTE interface remains in CR for a time $T_C$, waiting for new packets to be sent or received. If no packet is sent or received during $T_C$, the interface moves into the SHORT DRX state. In this state, the interface alternates sleep phases and short wake-up periods. The sleeping periods reduce the amount of power consumed by the interface, but during those periods, the interface is unable to receive incoming data. Wake-up periods are used to check for possibly incoming packets. Since the interface is not always ready to receive incoming transmissions, an additional delay may affect the reception of packets. If no packet is sent or received for a time equal to $T_S$, the interface transitions into the LONG DRX state. Similarly to SHORT DRX, in LONG DRX the LTE interface alternates sleeping periods with wake-up ones. However, in LONG DRX the sleeping periods are longer, trading increased delay during reception of packets for improved energy efficiency.
Finally, in the absence of transmissions and reception of packets, after a time $T_L$, the LTE module moves into the IDLE state. In this state, the interface sleeps most of the time, requiring the smallest amount of energy. Only when a new packet needs to be sent or received the interface returns into the CR state. However, the transition from the IDLE state to the CR state is not immediate, as a ``promotion'' period is needed to bring the interface into higher operational states. Note that the promotion period is not needed for the transitions from the SHORT DRX and the LONG DRX states to the CR state.

\subsection{Parameters of Operation}
\label{sec:parametersofoperation}

\begin{table}[]
\centering
\caption{The operational parameters of the LTE interface.}
    \begin{tabular}{lr}
        \toprule
        \textbf{Power consumption} & \textbf{Value (mW)} \\
        \midrule
        $P_{TX}$       & 1200       \tabularnewline
        $P_{RX}$          & 1000      \tabularnewline
        $P_C$           & 1000      \tabularnewline
        $P_S$               & 359.07    \tabularnewline
        $P_L$                & 163.23    \tabularnewline
        $P_I$                    & 14.25     \tabularnewline 
        $P_{PROM}$               & 1200      \tabularnewline 
        \bottomrule
    \end{tabular}
    
    \vspace{0.25cm}
  
    \begin{tabular}{lr}
        \toprule
        \textbf{Transition times} & \textbf{Duration (ms)} \\
        \midrule
        
        $T_C$   & 200       \tabularnewline
        $T_S$       & 400       \tabularnewline
        $T_L$        & 11000     \tabularnewline
        $T_{PROM}$       & 200       \tabularnewline 
        \bottomrule
    \end{tabular}
    
    \label{table:Parameters}
\end{table}

As previously stated, the operational parameters of the LTE interface that we use are the ones provided in \cite{10.1145/2796314.2745875}. The considered interface has $T_C$, $T_S$, and $T_L$ equal to 200~ms, 400~ms, and 11000~ms, respectively. 
The power consumed during the sending and the receiving phases depends on the quality of the Reference Signal Received Power (RSRP). We consider a power consumption $P_{TX}$ of 1200~mW for the sending phase and a power consumption $P_{RX}$ of 1000~mW for the receiving phase. In CR, outside any transmission period, a power consumption $P_C$ of 1000~mW is considered.
In SHORT DRX, the interface consumes 788~mW during a wake-up time, the duration of a wake-up time is 41~ms, and the interface wakes up with a period of 100~ms. In LONG DRX, a wake-up time consumes 788~mW for 45~ms and the interface wakes up with a period of 320~ms. During the sleeping phases, the energy consumption is equal to 61~mW (for both SHORT and LONG DRX).
The wake-up time in IDLE requires 570~mW with a duration of 32~ms, and the period is 1280~ms. The sleep phase in the IDLE state has negligible energy consumption. For simplicity, in SHORT DRX, LONG DRX, and IDLE we calculated the mean power consumption hereafter indicated as $P_S$, $P_L$, and $P_I$. Their values are equal to 359.07~mW, 163.23~mW, and 14.25~mW, respectively.
Finally, the LTE promotion requires an average power $P_{PROM}$ equal to 1200~mW, and its duration $T_{PROM}$ is equal to 200~ms. The parameters of the LTE interface are summarized in Table \ref{table:Parameters}.

\section{Evaluating the consumption of a request-response communication scheme}
\label{sec:consumption}

The energy $E_I$ needed by the LTE interface during a single iteration of the request-response protocol can be computed as
\begin{equation*}
    E_I = E_{TX} + E_W + E_{RX} + E_Q + E_{PROM_{TX}} + E_{PROM_{RX}} 
\end{equation*}
\noindent where $E_{TX}$ is the energy needed to send the data to the server, $E_W$ is the energy spent to wait for a response from the server, $E_{RX}$ is the energy needed to receive the response, $E_Q$ is the energy spent to wait for the beginning of the next iteration, and  $E_{PROM_{TX}}$ and $E_{PROM_{RX}}$ are the energy required to bring the LTE interface into CR before $T_{TX}$ and $T_{RX}$ respectively. 
The energy $E_{TX}$ can be computed as 
\begin{equation*}
    E_{TX} = T_{TX} \cdot P_{TX}
\end{equation*}
\noindent Similarly $E_{RX}$ can be computed as
\begin{equation*}
    E_{RX} = T_{RX} \cdot P_{RX}
\end{equation*}
\noindent Note that during $T_{TX}$ and $T_{RX}$, the interface remains in CR since packets are continuously sent and received (e.g. outgoing TCP segments and incoming ACKs). Instead, when the interface receives no packets, e.g. during $T_W$ and $T_Q$, it may move towards the states characterized by lower energy consumption. 

At the beginning of $T_W$, the LTE module is in CR, and it will remain in this state for a time $T_C$ before moving to SHORT DRX. This means that the interface will never move from CR if $T_W$ is smaller than $T_C$. In this case, the energy $E_W$ is equal to
\begin{equation*}
    E_W = T_W \cdot P_C 
\end{equation*}
Instead, if $T_W$ is greater than $T_C$ the interface stays in CR for $T_C$, and then it enters in SHORT DRX for the residual time. If the residual time is smaller than $T_S$ the interface remains in SHORT DRX and the energy consumption can be computed as 
\begin{equation*}
    E_W = T_C \cdot P_C + (T_W - T_C) \cdot P_S
\end{equation*}
If $T_W$ is larger, the interface remains in SHORT DRX for $T_S$, and then it goes into LONG DRX for the residual time. If the residual time is shorter than $T_L$, then $E_W$ can be expressed as 
\begin{align*}
    E_W &= T_C \cdot P_C + T_S \cdot P_S +(T_W - T_C -T_S) \cdot P_L
\end{align*}
Finally, when $T_W$ is even larger, the interface stays in LONG DRX for $T_L$ and then goes into the IDLE state for the remaining time, and in this case $E_W$ can be expressed as 
\begin{align*}
    E_W &= T_C \cdot P_C + T_S \cdot P_S + T_L \cdot P_L \\
        &+ (T_W - T_C - T_S -T_L) \cdot P_I
\end{align*}

To summarize, $E_W$ can be expressed as follows:
\[
\resizebox{.99\hsize}{!}{$
    E_W = 
    \begin{cases}
        T_W \cdot P_C
        &  \text{if } T_W \leq T_C\\
        
        T_C \cdot P_C + (T_W - T_C) \cdot P_S
        & \text{if } T_C < T_W \leq (T_C + \\
        & \quad T_S)\\
        
        T_C \cdot P_C + T_S \cdot P_S +   &  \text{if } (T_C + T_S) < T_W \leq (T_C + \\
        \quad (T_W - T_C - T_S) \cdot P_L & \quad T_S + T_L)\\
        
        T_C \cdot P_C + T_S \cdot P_S + T_L \cdot P_L   &  \text{if } (T_C + T_S + T_L) < T_W \\
        \quad + (T_W - T_C - T_S - T_L) \cdot P_I & \\
    
    \end{cases}
    $}
\]

$T_Q$ can be expressed as the difference between $T_I$, the period of operation of the application, and the time needed for a request-response phase:
\begin{equation*}
    T_Q = T_I - T_{TX} - T_{RX} - T_W - T_{PROM_{TX}} -  T_{PROM_{RX}}
\end{equation*}
\noindent where $T_{PROM_{TX}}$ and $T_{PROM_{RX}}$ are the time required to bring back the LTE interface into the sending/receiving states, if the LTE interface reaches the IDLE state. 
Since no packets are sent or received in $T_Q$, the energy spent during this phase can be computed similarly to $E_W$. Thus $E_Q$ can be expressed as:
\[
\resizebox{.99\hsize}{!}{$
    E_Q = 
    \begin{cases}
        T_Q \cdot P_C
        &  \text{if } T_Q \leq T_C\\
        
        T_C \cdot P_C + (T_Q - T_C) \cdot P_S
        & \text{if } T_C < T_Q \leq (T_C + \\
        & \quad T_S)\\
        
        T_C \cdot P_C + T_S \cdot P_S +   &  \text{if } (T_C + T_S) < T_Q \leq (T_C + \\
        \quad (T_Q - T_C -T_S) \cdot P_L & \quad T_S + T_L)\\
        
        T_C \cdot P_C + T_S \cdot P_S + T_L \cdot P_L   &  \text{if } (T_C + T_S + T_L) < T_Q \\
        \quad + (T_Q - T_C - T_S - T_L) \cdot P_I & \\
    
    \end{cases}
    $}
\]

Since the LTE promotion power $P_{PROM}$ is the same for transmission and reception, as well as the promotion duration $T_{PROM}$ we can state that:
\begin{equation*}
    E_{PROM_{TX}} = E_{PROM_{RX}} = E_{PROM} = T_{PROM} \cdot P_{PROM}
\end{equation*}
\noindent It has to be noted that $E_{PROM_{TX}}$ is consumed only if the LTE interface enters the IDLE state during $T_Q$, thus if $T_Q > (T_C + T_S + T_L)$. Similarly, $E_{PROM_{RX}}$ is consumed only if the LTE interface enters the IDLE state during $T_W$, thus if $T_W > (T_C + T_S + T_L)$.

From a different perspective, $T_W$ can be rewritten as
\begin{equation*}
    T_W = T_{ELAB} + RTT
\end{equation*}
\noindent where $RTT$ is the Round Trip Time (RTT) between the client and the server. The RTT value is going to be significantly smaller when the server is executed on the edge compared to when the server is executed in a possibly distant cloud infrastructure. Thus, the energy spent in $T_W$ depends on the RTT. Similarly, also $T_Q$ depends on the RTT, but in this case according to an opposite relationship: the larger the RTT, the smaller $T_Q$ (as mentioned, we assume that the application operates according to a fixed period $T_I$).

\section{Evaluation of Connectionless Communication}
\label{sec:udpresults}

We compared the energy consumption $E_I$ when the client interacts with both a server running on an edge node (hereafter $E_I^{E}$) and a server running in the cloud (hereafter $E_I^{C}$). In accordance with the scheme depicted in Figure~\ref{fig:Application}, we consider a simple application where the client uses UDP datagrams to exchange data with a server periodically. Since UDP lacks any rate control mechanisms (such as congestion and flow control mechanisms), the time required to send data depends only on the number of bytes transferred and the interface bitrate. Then, ${T_{TX}}$ can be computed as
\begin{equation*}
    T_{TX} = \frac{8 \cdot B_{TX}}{bitrate_{uplink}}
\end{equation*}
where $B_{TX}$ is the number of bytes sent by the client interface within a single period, including the overhead introduced by the underlying network stack levels.
We compute $T_{TX}$ and $T_{RX}$ considering an uplink bitrate of 1~Mbps and a downlink bitrate of 0.8~Mbps, respectively. With such low bandwidth bitrates, it is reasonable to assume that the bottleneck is located on the wireless segment of the path between C and S. As a consequence, we assume that packets received during ${T_{RX}}$ are received back-to-back on the client interface. Then, ${T_{RX}}$ can be computed as:
\begin{equation*}
    T_{RX} = \frac{8 \cdot B_{RX}}{bitrate_{downlink}}
\end{equation*}
where $B_{RX}$ is the number of bytes received on the client interface within a single iteration, including the overhead introduced by the underlying network stack levels. 

Results are presented by analyzing the ratio $\rho = E_I^{E} / E_I^{C}$. When $\rho$ is smaller than 1, the edge-based solution provides benefits in terms of energy consumption. The opposite occurs when $\rho$ is larger than 1. In our analysis, we assume a fixed value for the RTT towards the edge server, $RTT^{E}$, equal to 40 ms (this value is derived from a set of measurements described in \cite{CAIAZZA2021108140}). In the following sections, we analyze $\rho$ when varying the RTT towards the cloud $RTT^{C}$, the period of operation $T_I$, the amount of sent/received data, and the elaboration time $T_{ELAB}$, in different combinations.

\subsection{Varying the Period of Operation}
\label{subsection:Varying the application period}

\begin{figure}[!t]
    \centering
        
        \includegraphics[width=0.48\textwidth, trim={3.8cm 8.75cm 3.95cm 8.4cm}]{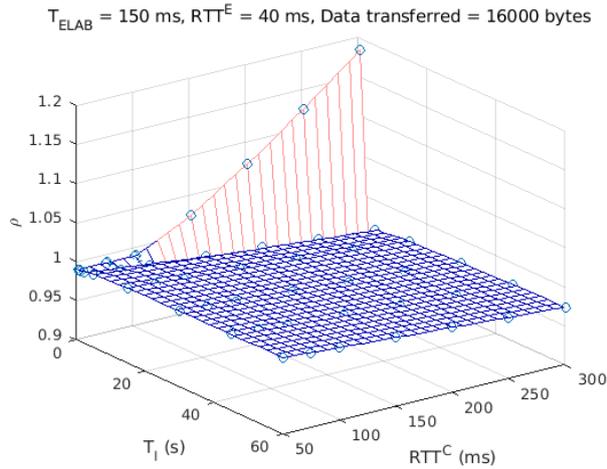}
    \caption{The ratio ($\rho$) between $E_I^E$ and $E_I^C$ for different values of $RTT^{C}$ when varying the period of operation $T_I$.}
    \label{fig:ratioincreasingTi_3D}
\end{figure}

\begin{table*}[!t]
    \centering
    \caption{$E_W$, $E_Q$, and $E_I$ values for both edge and cloud configurations in a connectioless communication scenario, for $T_I$ values of 750 ms and 1000 ms, and $RTT^{c}$ values ranging from 50 ms to 300 ms. The values have been computed considering $T_{ELAB}$ equals to 150 ms, $RTT^{e}$ equals to 40 ms, and an amount of data transferred equal to 16000 B for both $B_{TX}$ and $B_{RX}$.}
    \begin{tabular}{rrrrrrrr}
        \toprule
        $\bm{T_I}$ \textbf{(ms)} & $\bm{RTT^{C} (ms)}$ \textbf{(ms)} & $\bm{E_{W}^{E}}$ \textbf{(mJ)} & $\bm{E_{W}^{C}}$ \textbf{(mJ)} & $\bm{E_{Q}^{E}}$ \textbf{(mJ)} & $\bm{E_{Q}^{C}}$ \textbf{(mJ)} & $\bm{E_{I}^{E}}$ \textbf{(mJ)} & $\bm{E_{I}^{C}}$ \textbf{(mJ)} \tabularnewline 
        \noalign{\vskip 0.5mm}    
        \midrule
        \noalign{\vskip 0.5mm}  
        
        750&    50&     190.0&    200.0&    225.9&    222.3&
        \textcolor{tablegreen}{729.5}&
        \textcolor{tablered}{735.9}\tabularnewline
        
        750&    75&     190.0&    209.0&    225.9&    213.3&
        \textcolor{tablegreen}{729.5}&
        \textcolor{tablered}{735.9}\tabularnewline
        
        750&    100&    190.0&    218.0&    225.9&    204.3&
        \textcolor{tablegreen}{729.5}&
        \textcolor{tablered}{735.9}\tabularnewline
        
        750&    150&    190.0&    235.9&    225.9&    162.0& 
        \textcolor{tablered}{729.5}&  \textcolor{tablegreen}{711.5}\tabularnewline
        
        750&    200&    190.0&    253.9&    225.9&    112.0&
        \textcolor{tablered}{729.5}&  
        \textcolor{tablegreen}{679.5}\tabularnewline
        
        750&    250&    190.0&    271.8&    225.9&    62.0&
        \textcolor{tablered}{729.5}&  
        \textcolor{tablegreen}{647.4}\tabularnewline
        
        750&    300&    190.0&    289.8&    225.9&    12.0&
        \textcolor{tablered}{729.5}&  
        \textcolor{tablegreen}{615.4}\tabularnewline
        \midrule

        1000&      50&    190.0&  200.0&  315.6&  312.0&
        \textcolor{tablegreen}{819.2}& 
        \textcolor{tablered}{825.6}\tabularnewline
        
        1000&      75&    190.0&  209.0&  315.6&  303.1&
        \textcolor{tablegreen}{819.2}&  
        \textcolor{tablered}{825.6}\tabularnewline
        
        1000&     100&    190.0&  218.0&  315.6&  294.1&
        \textcolor{tablegreen}{819.2}&
        \textcolor{tablered}{825.6}\tabularnewline
        
        1000&     150&    190.0&  235.9&  315.6&  276.1&
        \textcolor{tablegreen}{819.2}&
        \textcolor{tablered}{825.6}\tabularnewline
        
        1000&     200&    190.0&  253.9&  315.6&  258.2&
        \textcolor{tablegreen}{819.2}&
        \textcolor{tablered}{825.6}\tabularnewline
        
        1000&     250&    190.0&  271.8&  315.6&  240.2&
        \textcolor{tablegreen}{819.2}&
        \textcolor{tablered}{825.6}\tabularnewline
        
        1000&     300&    190.0&  289.8&  315.6&  222.3&
        \textcolor{tablegreen}{819.2}&   \textcolor{tablered}{825.6}\tabularnewline
        \bottomrule
     
    \end{tabular}
    \label{table:Tiincreases_values}
\end{table*}

First of all, we evaluated the impact of the period of operation of the application on the terminal node. We assume a fixed amount of sent/received data of 16000~B and a fixed elaboration time $T_{ELAB}$ equal to 150~ms. Figure~\ref{fig:ratioincreasingTi_3D} shows $\rho$ when varying the application period ($T_I$), and the RTT towards the cloud server ($RTT^{C}$). The red region of the 3D surface corresponds to $\rho > 1$ values. This indicates that, for the combinations of $T_I$ and $RTT^{C}$ comprised in the red region, the interaction with a remote cloud requires less energy than the edge-based solution. Conversely, the blue region of the curve corresponds to the space of parameters where the edge-based approach is less energy-demanding. As can be observed, when $T_I$ is very small, the amount of energy needed to communicate with a cloud server is lower than the one needed to communicate with the edge server, especially when high $RTT^{C}$ values are considered. Instead, when $T_I$ becomes larger, the edge configuration becomes more energy efficient than the other one. 

Let's first understand why, for very small values of $T_I$, the cloud-based solution is more energy efficient. It has to be noted that $E_{TX}$ and $E_{RX}$ are the same in both edge and cloud configurations. This happens because these values only depend from the transmission and reception bitrate, which we assumed fixed, and the amount of data transferred, which we also assumed fixed. Thus, the only sources of differences between the edge and the cloud configurations are $E_W$ and $E_Q$, which are influenced by the RTT. The $T_W$ for the edge configuration (hereafter $T_W^E$) is equal to 190 ms: 150 ms for $T_{ELAB}$, and 40 ms for the RTT towards the edge (hereafter $RTT^{E}$). With this duration, the LTE interface remains in the CR state for the entire duration of $T_W^E$, as we recall that the LTE interface transitions from the CR state to the SHORT DRX state after 200~ms without sending or receiving a packet. In the cloud configuration instead, $RTT^{C}$ is higher than $RTT^{E}$, thus, for fixed $T_I$ values, $T_W^C$ is higher than $T_W^E$ and $T_Q^C$ is smaller than $T_Q^E$, especially for high $RTT^{C}$ values. Let $\Delta RTT$ be the difference between the cloud and the edge RTT:
\begin{equation*}
    \Delta RTT = RTT^{C} - RTT^{E}
\end{equation*}
\noindent Consequently
\begin{equation*}
\begin{split}
    T_{W}^{C} = T_{W}^{E} + \Delta RTT\\
    T_{Q}^{C} = T_{Q}^{E} - \Delta RTT
\end{split}
\end{equation*}
Differently from the edge configuration, in the cloud configuration, during $T_W^C$, the LTE interface stays in the CR state for 10~ms more, then it transitions to the SHORT DRX state. This means that, in the cloud configuration, the LTE interface spends most of the additional $\Delta RTT$ of $T_W^C$ in a state characterized by relatively low power consumption. Thus the additional amount of energy consumed in $T_W^C$, with respect to the energy consumed in $T_W^E$ is not so high. 
For small values of $T_I$ and high $RTT^{C}$, $T_{Q}^{C}$ lasts only few milliseconds, which, even if LTE interface spends them in the CR state, result in a low amount of spent energy. Instead, in the edge configuration, the LTE interface spends most of the additional $\Delta RTT$ of $T_{Q}^{E}$ in the CR state, with a high power consumption.
This means that the additional energy spent by the cloud configuration in $T_{W}^{C}$ is lower than the additional energy spent by the edge configuration in $T_{Q}^{E}$. As an example, Table~\ref{table:Tiincreases_values} shows the values of $E_W$, $E_Q$, and $E_I$ for both the edge and cloud configuration, when varying $RTT^{C}$, for two possible values of $T_I$, 750 ms and 1000 ms, for illustrating the source of differences. For simplicity, for the edge and the cloud configurations we added to all the parameters the superscripts ``$E$'' and ``$C$'', respectively. This notation will be kept for the rest of the paper. As can be observed, when $T_I$ is equal to 750 ms, for higher $RTT^{C}$ values, $E_Q^{C}$ drops down to 12.0 mJ, and $E_I^{C}$ becomes smaller than $E_I^{E}$.

Let's now analyze why, for larger $T_I$ values, the edge-based configuration is more energy efficient (the blue area in Figure~\ref{fig:ratioincreasingTi_3D}). When $T_I$ increases there is no impact on $T_W$, thus $E_W$ remains smaller for the edge configuration. Conversely, in both edge and cloud configurations, $T_Q$ and $E_Q$ increase as the period of operation gets larger. When $T_I$ is equal to 1~s, in the cloud configuration, the residual time $T_Q^C$ is barely sufficient to let the LTE interface enter the SHORT DRX state. This means that for larger $T_I$ the energy spent in $T_Q$ becomes very close for the two configurations and $E_W$ is the component that makes the difference.

\subsection{Varying the Amount of Sent/Received Data}

\begin{figure}[!t]
    \centering
        \includegraphics[width=0.47\textwidth,  trim={3.7cm 8.55cm 4.2cm 8.5cm}]{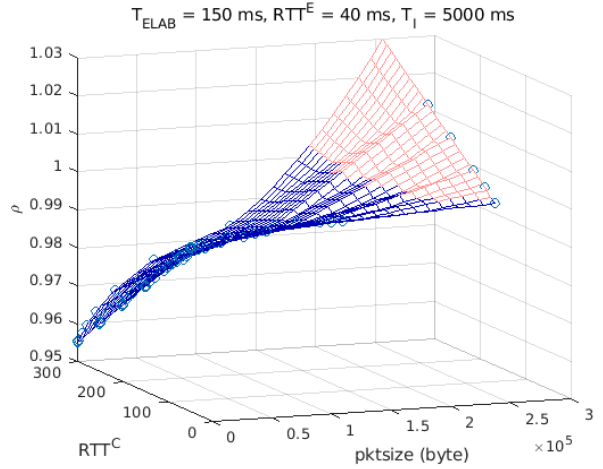}
    
    \caption{The ratio ($\rho$) between $E_I^E$ and $E_I^C$ for different values of $RTT^{C}$ when varying the amount of transferred data.}
    \label{fig:ratioincreasingpktsize_3D}
\end{figure}

We now evaluate how the amount of data transferred during requests and responses impacts on the energy needed for the two configurations. Figure~\ref{fig:ratioincreasingpktsize_3D} shows the value of $\rho$ when the period of operation is fixed ($T_I$ = 5000~ms), the elaboration time is fixed ($T_{ELAB}$ = 150~ms), the amount of data sent/received during each request/response varies from few bytes to 256~KB, and the cloud RTT, $RTT^{C}$, varies from 50~ms to 300~ms.
When the amount of transferred data is small, the edge-based configuration is always more energy efficient, especially when $RTT^{C}$ is high. The reasons for this behavior are similar to the ones described in Section~\ref{subsection:Varying the application period}. Since the $RTT^C$ is higher, also $T_W^C$ is higher. A fraction of the additional time $\Delta RTT$ is spent in the CR state, with high power consumption. When the amount of data transferred is small, $T_{TX}$ and $T_{RX}$ are small for both edge and cloud, while $T_Q^E$ and $T_Q^C$ are big enough to let the interface enter the LONG DRX state for both the configurations. This means that the parameter that most influences the energy consumption is $T_W$, which strongly depends on the RTT. Since the $RTT^C$ is higher, the energy consumption is higher in cloud configurations.

On the contrary, when a higher amount of data is sent/received, $T_{TX}$ and $T_{RX}$ are larger and, as a consequence, $T_Q^E$ and $T_Q^C$ become smaller. When $T_Q^E$ and $T_Q^C$ become so small that the LTE interface is unable to move to the lower energy states (e.g., LONG DRX), then the global consumption becomes more favorable to the cloud-based configuration.

\subsection{Varying the Elaboration Time} \label{subsection:Varying the elaboration time}

\begin{figure}[!t]
    \centering
        \includegraphics[width=0.48\textwidth, trim={3.7cm 8.6cm 3.55cm 8.5cm}]{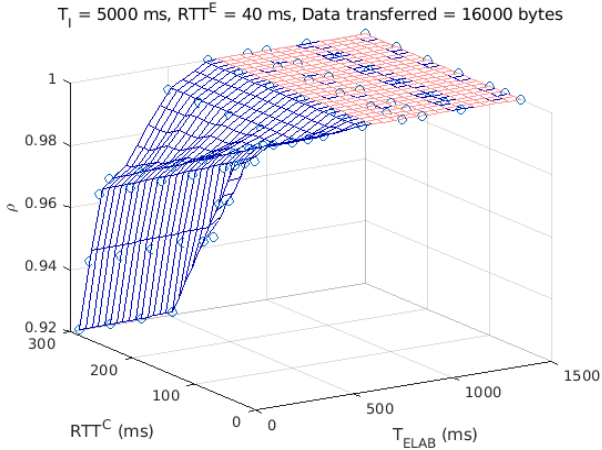}
    \caption{The ratio ($\rho$) between $E_I^E$ and $E_I^C$ for different values of $RTT^{C}$ when varying elaboration time $T_{ELAB}$.}
    \label{fig:ratioincreasingTelab_3D}
\end{figure}

We now evaluate the impact of the elaboration time of the server. The value of $\rho$ for different values of $T_{ELAB}$ and $RTT^{C}$ is shown in Figure~\ref{fig:ratioincreasingTelab_3D}. For this analysis, we set $T_I$ = 5000~ms and an amount of sent/received data of 16000~B. $T_{ELAB}$ is varied between 0~ms and 1500~ms, and $RTT^{C}$ between 50~ms and 300~ms.
The figure shows that the edge configuration is always convenient for small $T_{ELAB}$ values. This happens because small $T_{ELAB}$ values in the edge configuration produce small $T_W^E$ values, which the LTE interface spends in the CR state, but for a small amount of time. In the cloud configuration instead, especially when high $RTT^C$ are considered, $T_W^C$ is much longer than $T_W^E$, thus the amount of time spent in CR is higher. With $T_I$ = 5000~ms, both $T_Q^E$ and $T_Q^C$ are long enough to make the LTE interface stay in CR for the entire time, and move to less energy demanding states. When $T_{ELAB}$ is small, the total amount of time spent in CR in the cloud configuration is higher than in the edge configuration, and as a consequence also the energy spent is higher.

When $T_{ELAB}$ increases, both $T_{W}^E$ and $T_W^C$ increase accordingly. For a given $T_{ELAB}$ on, in the edge configuration the LTE interface is able to move into the SHORT DRX state or even in the LONG DRX state during $T_W$. This decreases the benefits obtained during $T_W^E$ by the edge-based solution. In fact, as can be seen, from a given $T_{ELAB}$ on, choosing either an edge or a cloud configuration does not make any difference from an energy consumption perspective.

\subsection{Varying the Elaboration Time and the Application Period}

\begin{figure}[!t]
    \centering
    \includegraphics[width=0.48\textwidth, trim={3.7cm 8.7cm 4.2cm 8.6cm}]{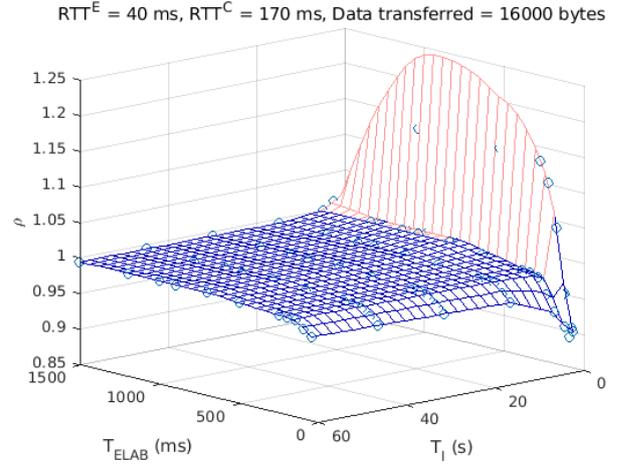}
    
    \caption{The ratio ($\rho$) between $E_I^E$ and $E_I^C$ when varying the elaboration time $T_{ELAB}$ and the period of operation $T_I$.}
    \label{fig:ratioincreasingTelabTi_3D}
\end{figure}

Finally, we evaluate the impact of varying the elaboration time $T_{ELAB}$ and the period $T_I$, while keeping unchanged the cloud RTT (i.e. $RTT^{C}$) to 170~ms, and the amount of transmitted data to 16000~B (Figure~\ref{fig:ratioincreasingTelabTi_3D}). The cloud-based configuration brings benefits only when $T_I$ is really small. For almost all the other configurations, there is almost no benefit in either the edge or the cloud solutions, except for extremely low values of $T_{ELAB}$. In this latter case, when the server is running in the edge, the energy consumption is lower. This happens because a short $T_{ELAB}$ can reduce the time spent in CR during $T_W$ for both edge and cloud configurations. When the server is in the cloud, in the $T_W^C$ interval, the LTE interface spends a relevant part of $\Delta RTT$ in CR, with significantly higher power consumption. Instead, when $T_{ELAB}$ becomes larger than 160~ms, the consumption during $T_W$ for both configurations becomes similar. In this scenario, the difference in energy consumption between the two configurations originates from $T_Q$. For small $T_I$ values, the cloud-based solution is more energy efficient, as the residual time $T_Q^C$ is smaller than the corresponding time of the edge-based solution, thus less energy demanding. 

\subsection{Impact of Communication Latency on Communication Costs: an Example}
\label{sec:example}

Let's suppose that a terminal node produces a given amount of data $X$ every hour. Let's also suppose that communication costs can be defined to take into account not only the energy expenditure of the terminal node, but also the delay that the data incur waiting to be transferred to the server side.  Communication cost $C$ could be defined as

\begin{equation*}
    C = \alpha*\frac{E}{E_{MAX}} + (1-\alpha)*\frac{D}{D_{MAX}}
\end{equation*}

\noindent where $E$ is the total energy needed to transfer the data, and $D$ is the time the data wait on the client side. The latency $D$ is basically the $T_I$ period considered so far. $\alpha$ is a coefficient that can be used to weigh the two components depending on the user's preference for saving energy or obtaining a reduced delay. $E_{MAX}$ is the maximum total energy needed for transferring the data and  $D_{MAX}$ is the maximum possible delay. Such maximum values depend on the considered space of parameters and are used to normalize the two cost components before assigning them a weight. Since the total amount of data $X$ in an hour must be always the same, the amount of data that must be included in a single request-response cycle is equal to $X / (3600~\text{s} / T_{I})$. The reply is supposed to be just 1 byte, to confirm the reception of data. 

\begin{figure}[!t]
    \centering
    \includegraphics[width=0.9\columnwidth]{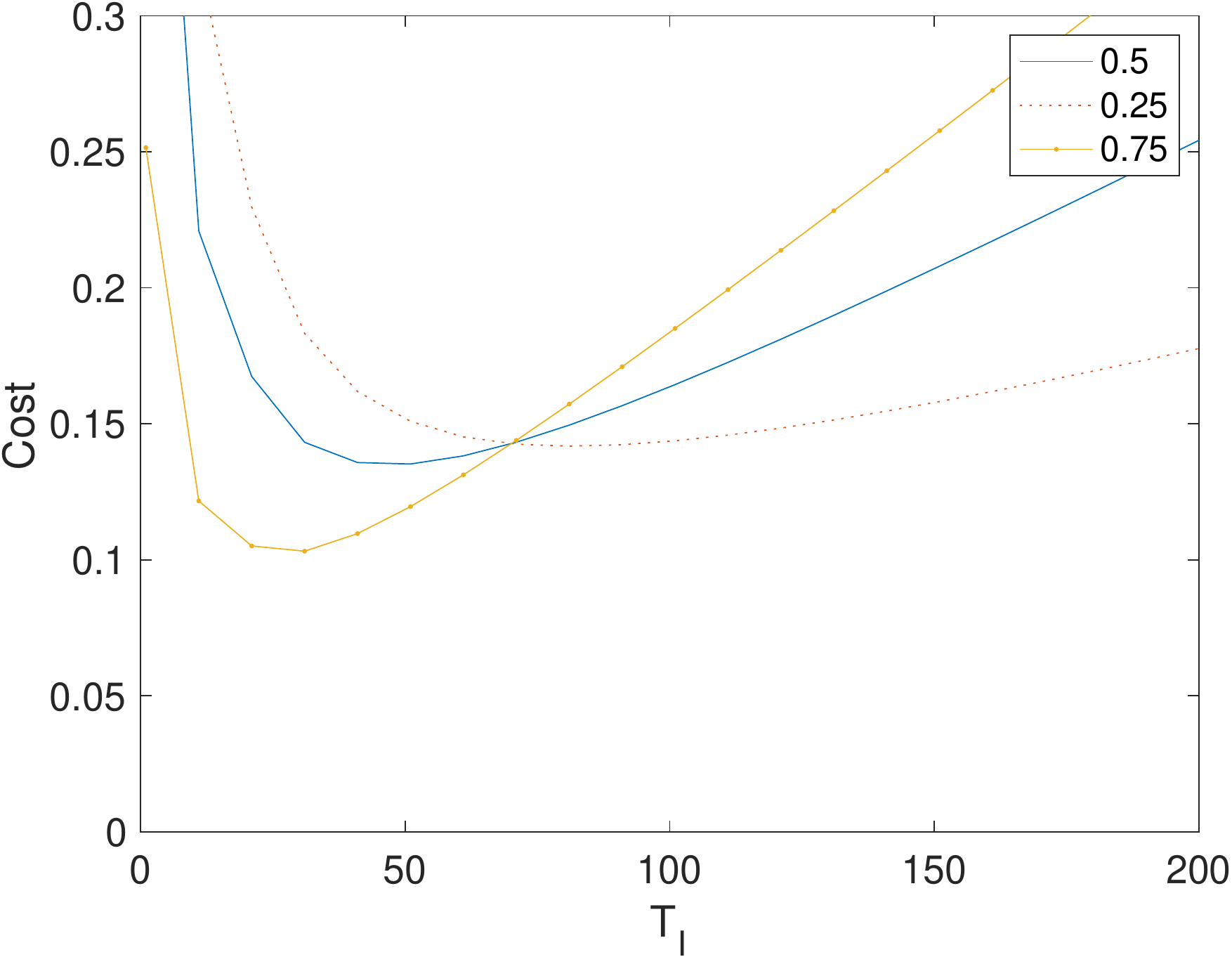}
    \caption{Communication costs.}
    \label{fig:opt}
\end{figure}

Figure~\ref{fig:opt} shows how $C$ changes when varying $T_I$. The latency between the terminal node and the server is supposed to be equal to 50 ms, and $X$ is equal to 10e6 B. The three curves correspond to values of $\alpha$ equal to 0.25, 0.5, and 0.75. The minimum cost is obtained when the $T_I$ period is approximately equal to 30, 60, and 80 s, depending on the weight given to the two components. 

\section{Evaluation of Connection-Oriented Communication}
\label{sec:tcpresults}

In this section, we consider the performance of a TCP-based application. When TCP is involved, the maximum throughput achievable at the steady-state is limited by multiple factors. For example, the flow control mechanism guarantees that the amount of data sent can not exceed the receiving windows. If a device with poor receiving capabilities is considered, this mechanism can affect the throughput. Instead, if the buffer at the receiver side is sufficient to handle incoming traffic, the steady-state throughput will be determined by the congestion control mechanism.

For different TCP congestion control strategies (e.g., Reno, Vegas, CUBIC), the steady-state throughput has been shown to be always dependent on the RTT~\cite{10.1145/263932.264023, 10.1145/285243.285291, cardwell1998modeling, Bao2010:model}, even if in different forms. The models proposed in the literature are however valid under precise assumptions and contain parameters (e.g., the packet loss probability) whose values can change over time and can only be collected through network measurements. For this reason, to overcome the limitations of an analytical analysis of TCP flows, we conducted a set of experiments aimed at obtaining the values of $T_{TX}$, $T_{RX}$, $T_W$, and $T_Q$ in a real LTE environment. These values are then included in our model to calculate $\rho$ for connection-oriented communications.

\subsection{Application Models}
 \begin{figure}[!t]
 \centering
     \subfloat[Uploading a file to the server.]{
        \includegraphics[width=0.22\textwidth]{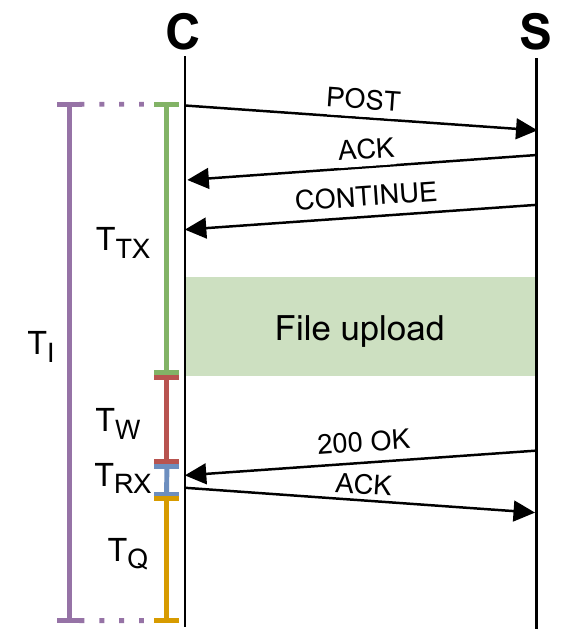}
        \label{fig:app1}
    }
    \subfloat[Downloading a file from the server.]{
        \includegraphics[width=0.23\textwidth]{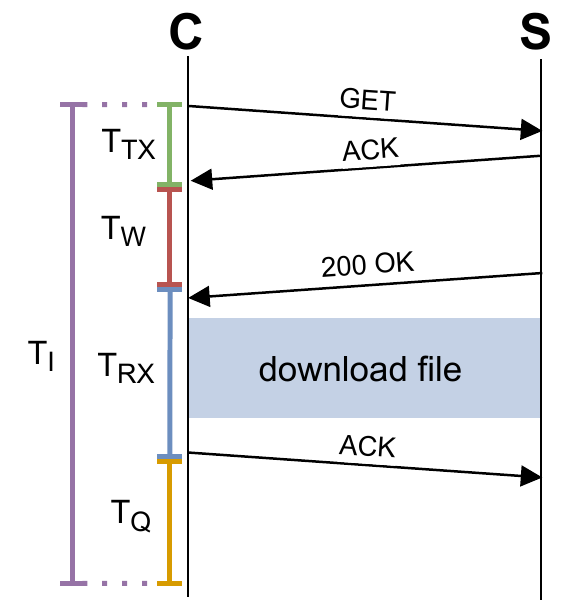}
        \label{fig:app2}
    }
    \caption{Connection-oriented application models.}
    \label{fig:httpapp}
\end{figure}

According to the scheme of Figure~\ref{fig:Application}, we consider two applications that periodically exchange data on an existing TCP connection. The first application consists of a client which sends a relatively large amount of data to a server and receives small responses. For example, this behavior can be representative of a smart camera that uploads footage data on a server and receives a confirmation message. The second application implements the opposite scenario, where a client sends small requests and receives larger responses. This second application can be representative of a device that periodically polls a centralized server and receives some updates, as many smart objects or mobile applications do. In practice, these behaviors are mapped on an HTTP POST request and an HTTP GET request, respectively (Figure~\ref{fig:httpapp}). In the first application, shown in Figure~\ref{fig:app1}, we consider $T_{TX}$ as the time that goes from the start of the HTTP POST request to the reception of the last ACK from the server. $T_{RX}$ is instead the time that goes from the reception of the HTTP response from the server to the sending of the ACK from the client. In the second application, shown in Figure~\ref{fig:app2}, we consider $T_{TX}$ as the time from the start of the HTTP GET request to the reception of the ACK from the server. $T_{RX}$ is instead the time required to receive the response from the server, which also includes the file transfer, and send the ACK back to the server.

For both applications, we still consider $T_W$ as the time elapsed between the end of $T_{TX}$ and the start of $T_{RX}$, and $T_Q$ as the residual time required to complete the current operational period $T_I$.

\subsection{Experimental Setup}

\begin{figure}[!t]
    \centering
    \includegraphics[width=0.35\textwidth]{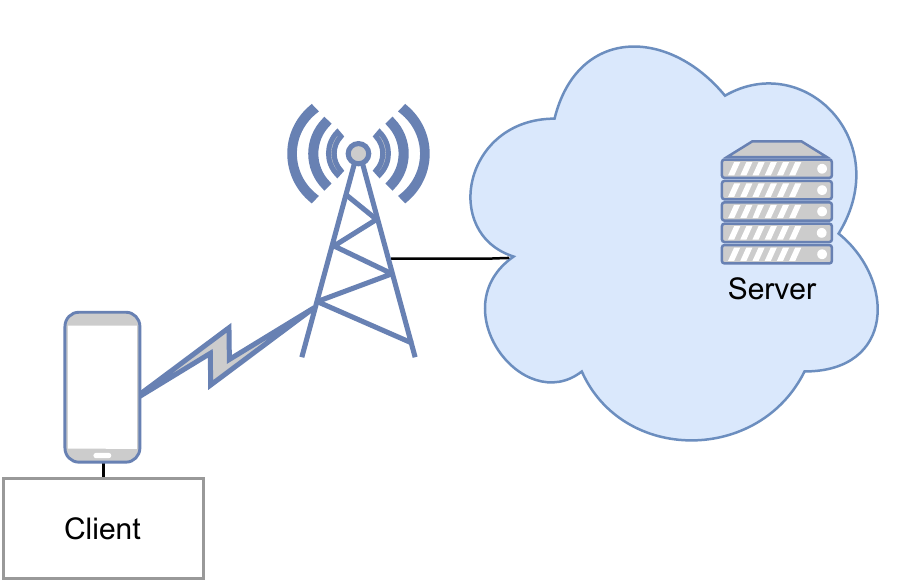}
    \caption{Experimental setup for the connection-oriented communication evaluation.}
    \label{fig:setup}
\end{figure}

To obtain the values of $T_{TX}$, $T_{RX}$, $T_W$, and $T_Q$ of both edge and cloud configurations in a realistic environment, we conducted a set of experiments using the setup of Figure~\ref{fig:setup}. We used a Raspberry Pi to host the client application. The Raspberry Pi was chosen to include in the experiments the possible effects caused by a device with limited computational capabilities, as these devices are the subject of this study. The Raspberry Pi is connected to an Android smartphone, which provides LTE connectivity through a USB connection.
The client application sends HTTP requests towards an NGINX server hosted at the University of Pisa for both edge and cloud experiments. We observed approximately 75~ms of average RTT, measured with ping, between the client and the server. Then, to mimic a cloud environment, we used \textit{tc} to add a 100~ms delay on the server interface for both incoming and outgoing traffic. This means that $\Delta RTT$ = 200~ms was added for the cloud configuration.

\begin{algorithm}
    \caption{The client algorithm executed on the RaspberryPi during the experimental phase.}
    \label{alg:connectionOrientedSetup}
    \For{i $\in$ \{1, 2, 3, 4, 5, 6, 7, 8, 9, 10\}}{
        \For{f $\in$ \{2, 4, 6, 8, 10, 12\} MB}{
            \For{s $\in$ [$server_{EDGE}$, $server_{CLOUD}$]}{
                Start the tshark session\;
                Send an HTTP POST request to upload on server s a file of size f\;
                Send an HTTP GET request to download from server s a file of size f\;
                Stop the tshark session and store the cap results\;
            }
        }
    }
\end{algorithm}

The experiments have been conducted as shown in the pseudo code of Algorithm \ref{alg:connectionOrientedSetup}.
At the beginning, the client starts a sniffing session between the client and the server using \textit{tshark}. 
At this point, the client uses \textit{curl} to send to the edge server two HTTP requests. The former is a POST request that implements the behavior of the first application, while the latter is an HTTP GET request that implements the behavior of the second application.  
Finally, the client stops the sniffing session, storing the collected trace inside a \textit{cap} file for further processing.
Then, the procedure is repeated using the cloud server as a target and keeping the size of the uploaded/downloaded file unchanged. Finally, once both edge and cloud metrics have been collected the whole procedure is repeated considering a larger file. 

We conducted two experiments. The first experiment was run overnight to collect  $T_{TX}$, $T_{RX}$, and $T_W$ values in a scenario where the cell is lightly loaded, while the second one was run during the day to collect metrics in the presence of a higher network load.
Finally, the values of $T_{TX}$, $T_{RX}$, and $T_W$ are extracted from the \textit{cap} file, and $T_Q$ is computed as the difference between $T_I$ and the sum of the other values. These values are then used to compute the corresponding energy values, as explained in Section~\ref{sec:consumption}. It must be noted that each HTTP request of the experimental phase was performed independently from the others using curl on different TCP connections, while the models considered for the two applications assume to operate on existing connections. In addition, we did not consider the packets exchanged to open and close the TCP connections.

\subsection{Results}

\begin{figure*}[!t]
 \centering
     \subfloat[Night time, HTTP POST application.]{
        \includegraphics[width=0.47\textwidth]{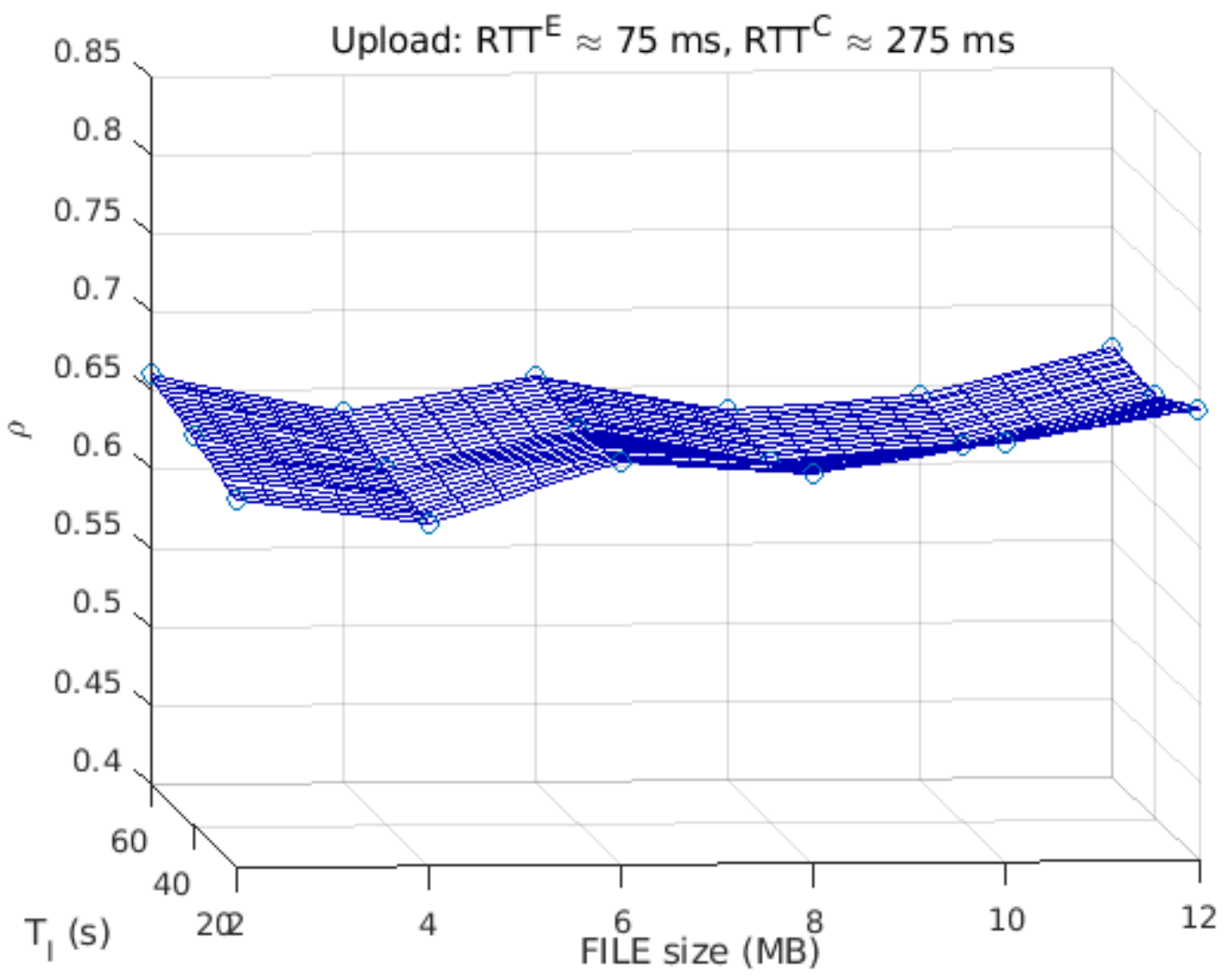}
        
        \label{fig:night-app1-results}
    }
    \subfloat[Night time, HTTP GET application.]{
        \includegraphics[width=0.47\textwidth]{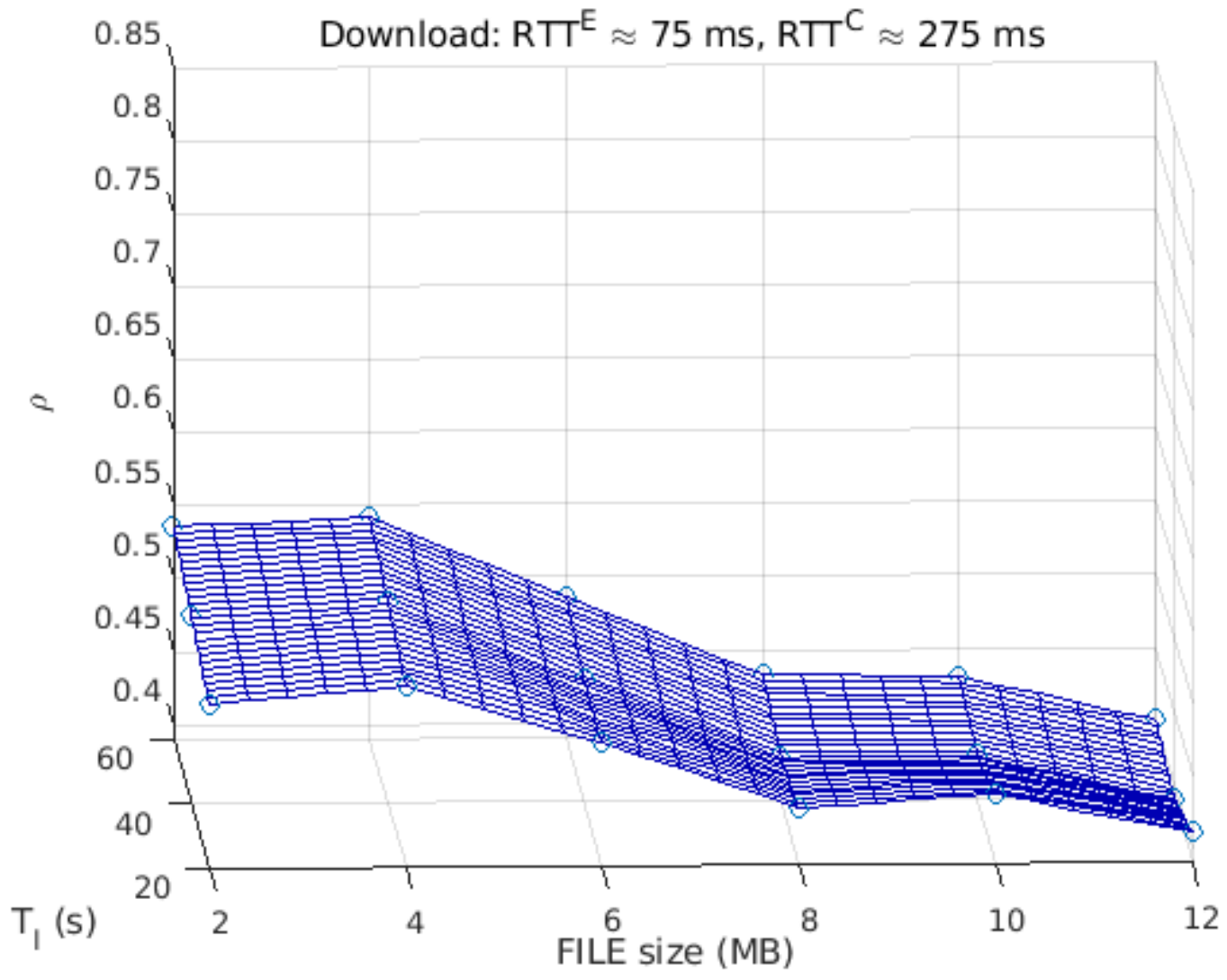}
        \label{fig:night-app2-results}
    }\\
     \subfloat[Day time, HTTP POST application.]{
        \includegraphics[width=0.47\textwidth]{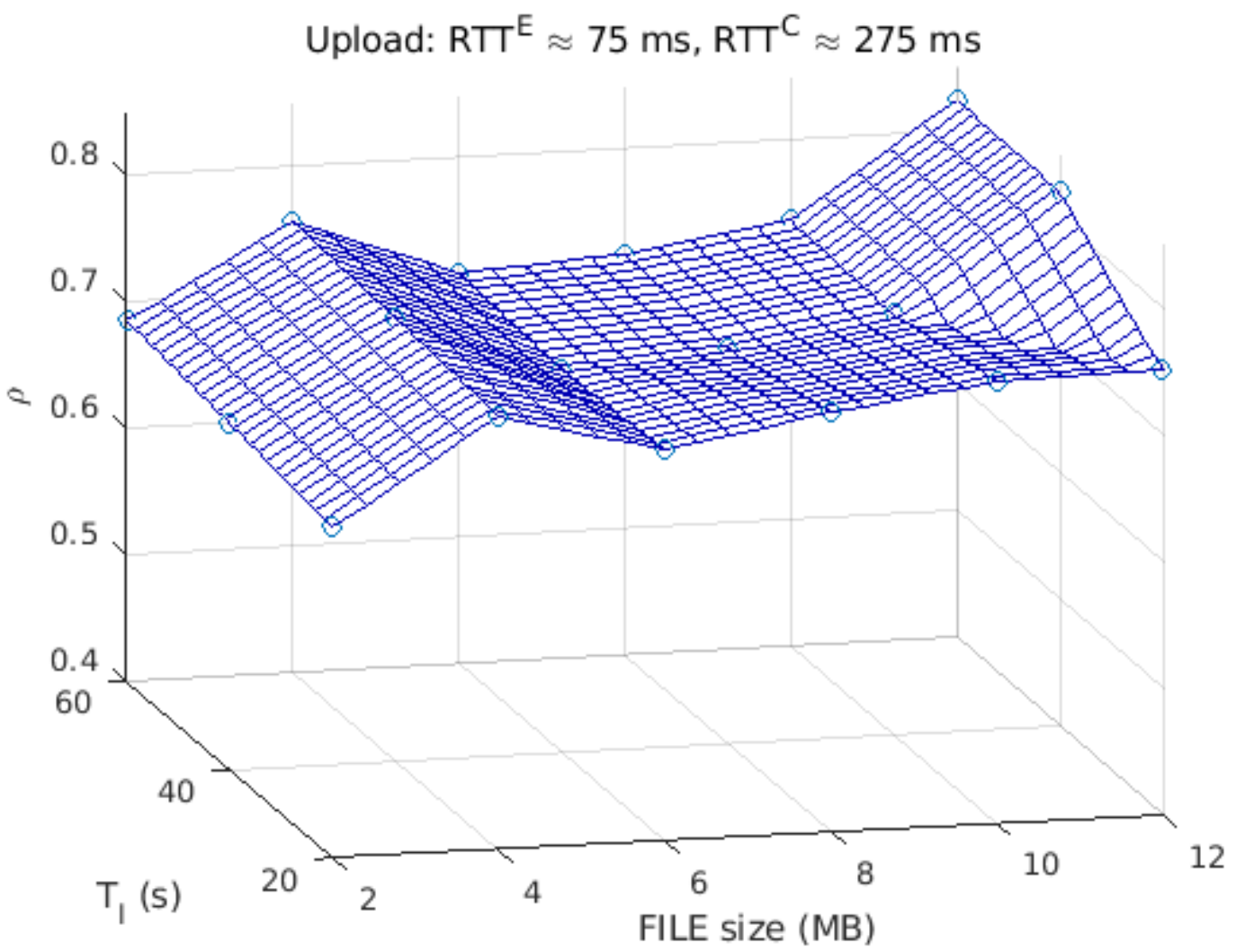}
        
        \label{fig:day-app1-results}
    }
    \subfloat[Day time, HTTP GET application.]{
        \includegraphics[width=0.47\textwidth]{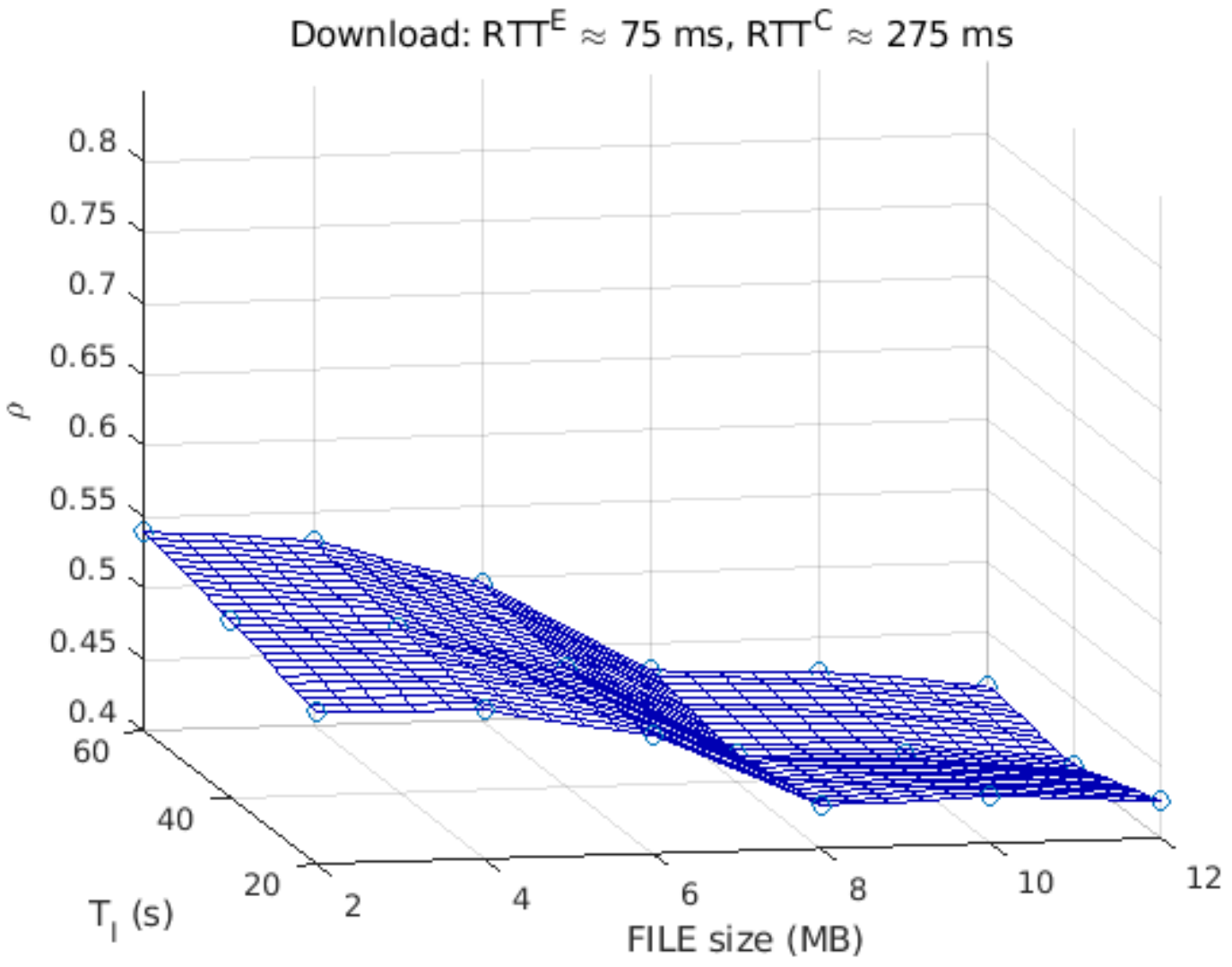}
        \label{fig:day-app2-results}
    }
    \caption{The ratio ($\rho$) between the $E_I^E$ and $E_I^C$, when varying the application period $T_I$ and the size of the transferred file.}
    \label{fig:realdata-results}
\end{figure*}

\begin{figure*}[!t]
 \centering
     \subfloat[Night time, HTTP POST application.]{
        \includegraphics[width=0.49\textwidth]{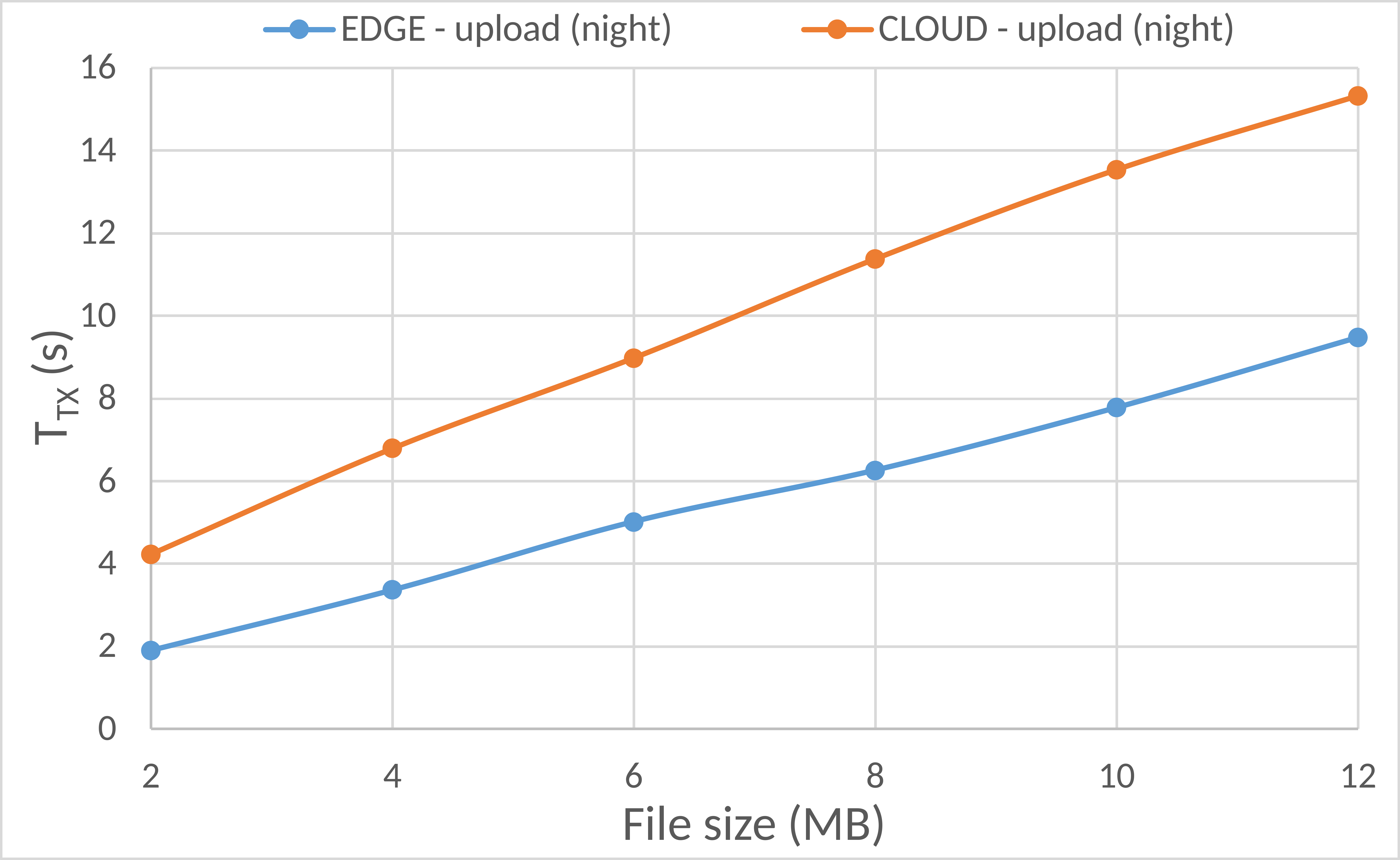}
        \label{fig:night-app1-time-results}
    }
    \subfloat[Night time, HTTP GET application.]{
        \includegraphics[width=0.49\textwidth]{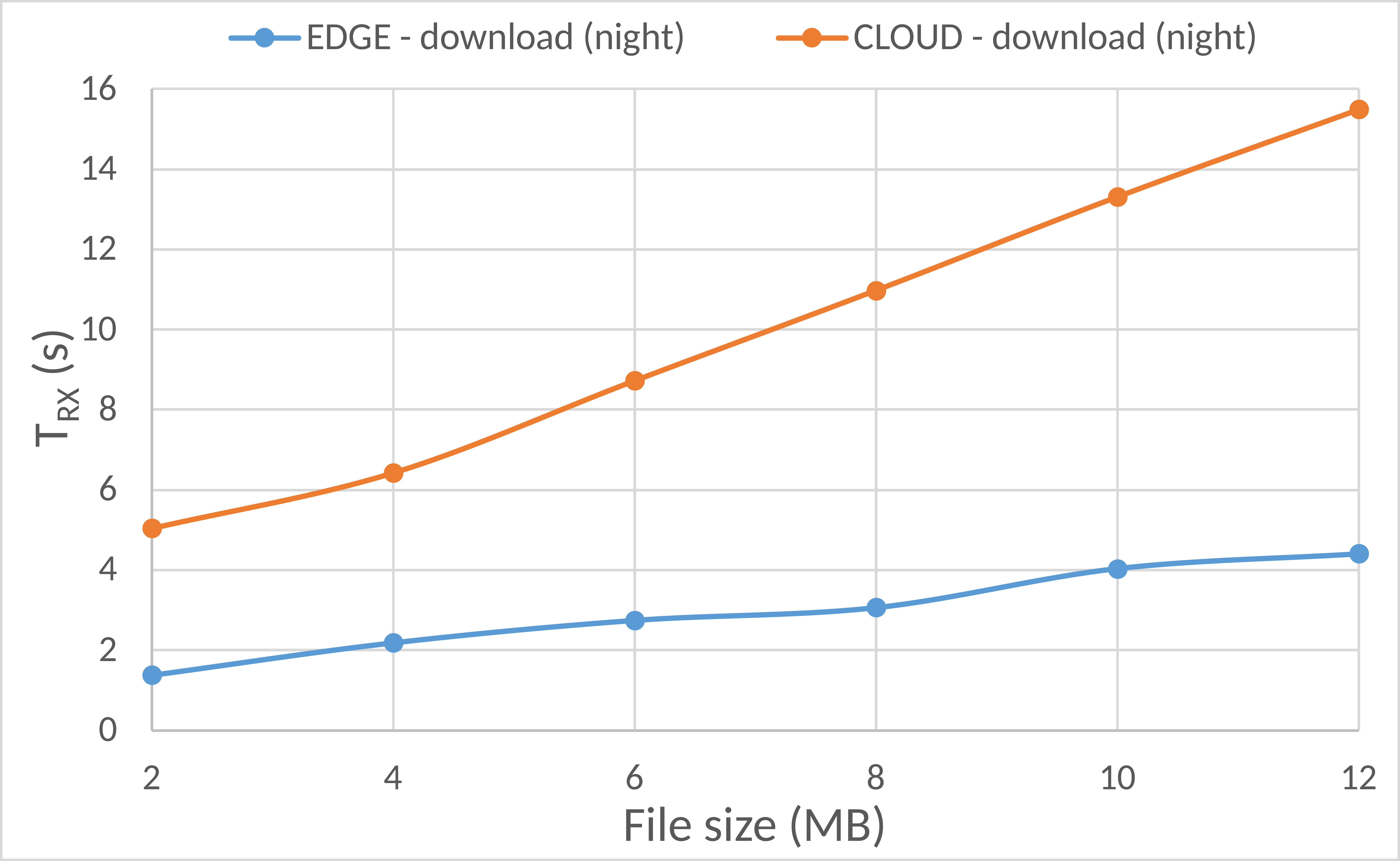}
        \label{fig:night-app2-time-results}
    }\\
    \subfloat[Day time, HTTP POST application.]{
        \includegraphics[width=0.49\textwidth]{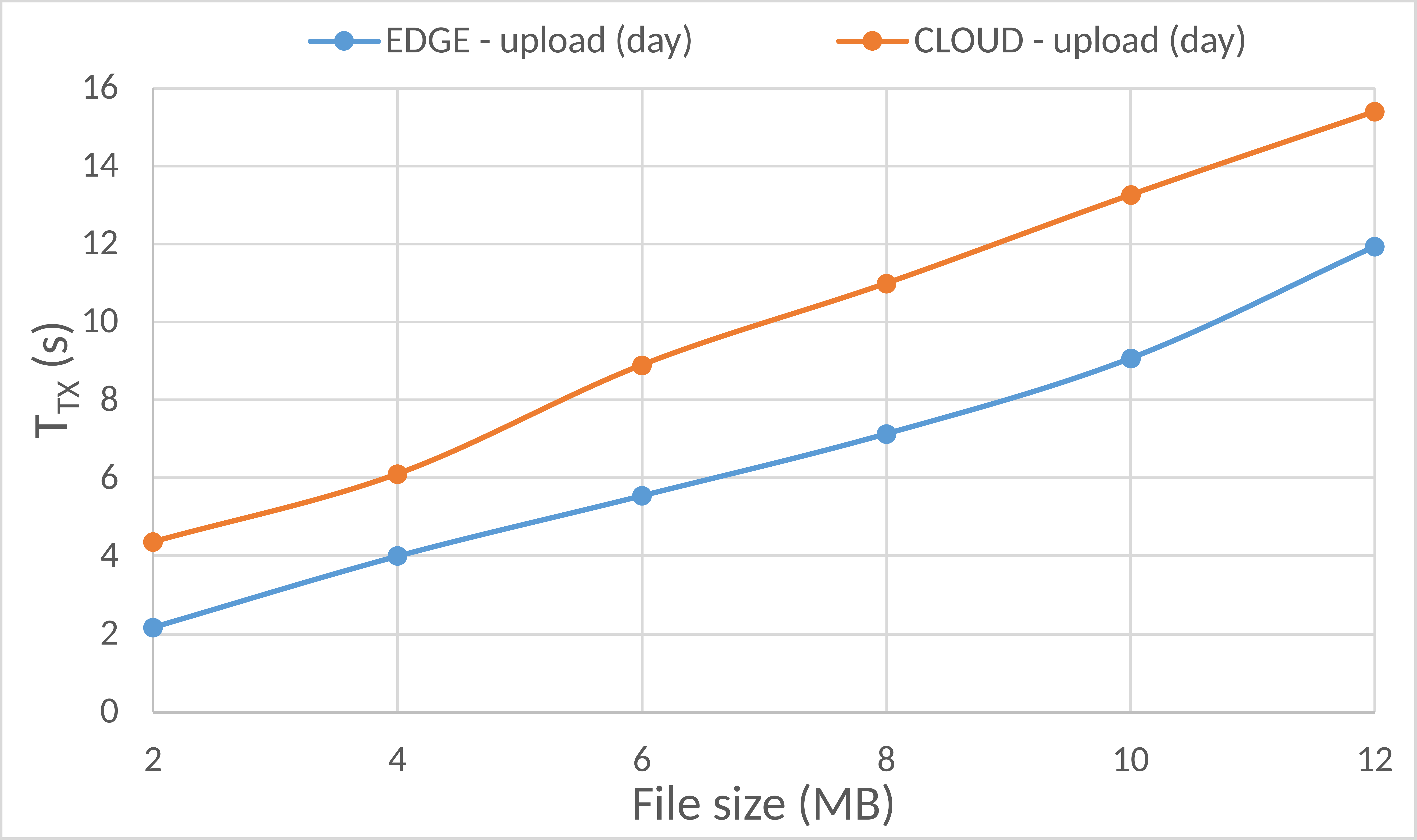}
        \label{fig:day-app1-time-results}
    }
    \subfloat[Day time, HTTP GET application.]{
        \includegraphics[width=0.49\textwidth]{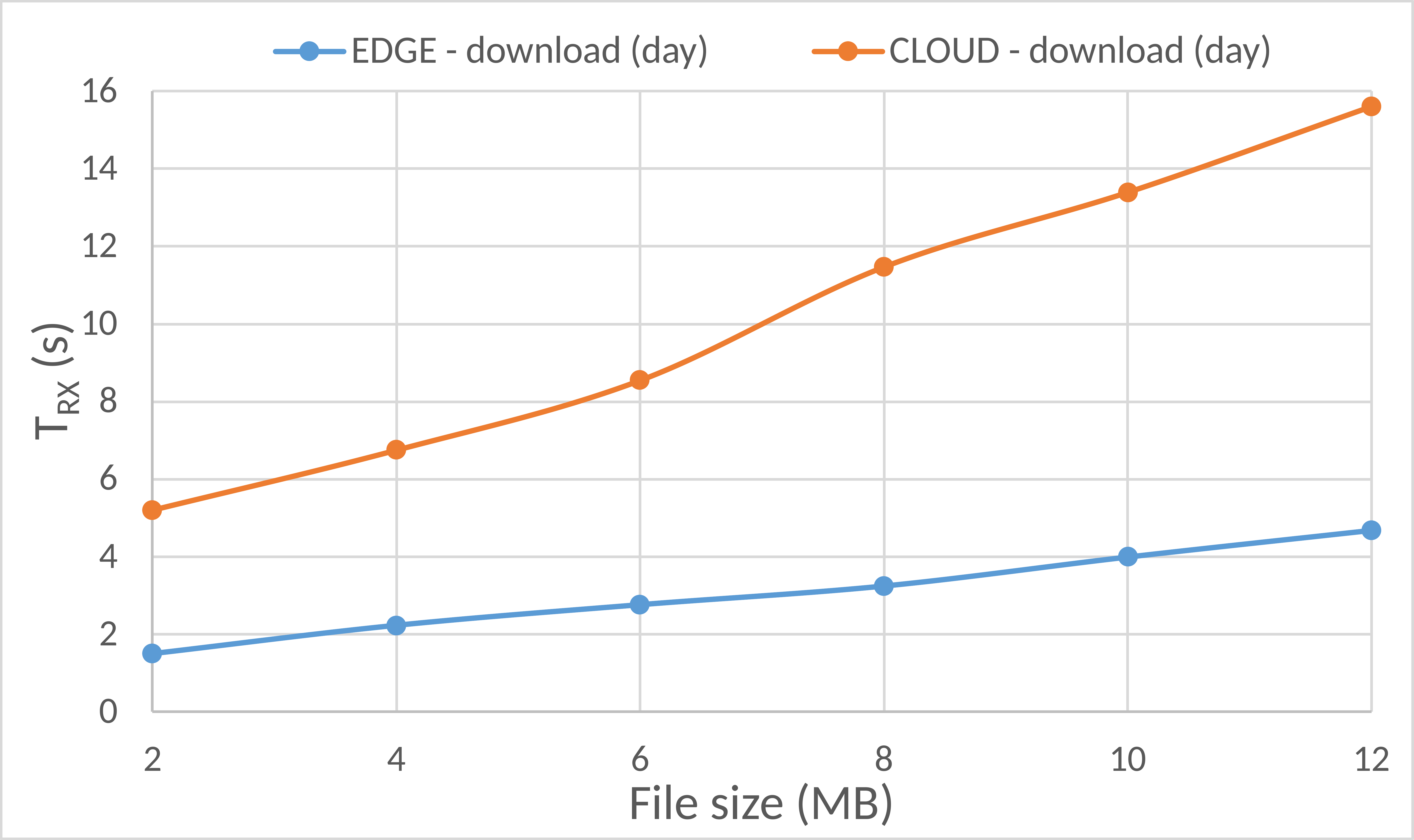}
        \label{fig:day-app2-time-results}
    }
    \caption{The mean time required to transmit data upon 10 repetitions.}
    \label{fig:realdata-time-results}
\end{figure*}

For each $T_I$ and each file size, we define $E_I^{E}$ and $E_I^{C}$ for a connection-oriented communication as the sum of the energy consumed over the 10 repetitions of each experiment involving the edge and the cloud server, respectively. We consider $T_I$ values from 20 seconds to 60 seconds. The ratio $\rho$ between $E_I^{E}$ and $E_I^{C}$ is depicted in Figure~\ref{fig:realdata-results} for the two applications, and night and day scenarios.

First, we can notice that the value of $\rho$ is always smaller than 1. This means that the edge configuration is always convenient as it has smaller energy consumption. This happens because, as stated above, the throughput of a TCP connection depends on the RTT. Thus, the throughput is higher, and increases faster, when smaller RTT are involved, i.e., when the server is placed on the edge node. This implies a shorter time to upload (or download) a file, and, as a consequence, a shorter amount of time spent in the CR state, the most energy consuming state, by the client LTE interface. Therefore, the energy consumption of the edge configuration is smaller. As a confirmation, in Figure~\ref{fig:realdata-time-results} we show the transmission times for uploading and downloading a file in both edge and cloud, and night and day scenarios. We can observe that the transmission times of the edge configuration are always lower than those of the cloud configuration. However, the difference between the transmission times is not always the same. For example, for the HTTP POST application, in both night time and day time, the transmission time obtained in the edge scenario grows with approximately the same slope as the transmission time obtained in the cloud scenario. For the HTTP GET application instead, the transmission times in the cloud scenario grow much faster than in the edge scenario. This reflects also in the different shapes of surfaces of Figure~\ref{fig:realdata-results}. For example, for the HTTP POST application, in the night time, the surface is basically flat (Figure~\ref{fig:night-app1-results}). In the day time instead, $\rho$ grows as the file size grows, especially for high $T_I$ times (Figure~\ref{fig:day-app1-results}). This happens because the transmission times of the edge and the cloud scenarios in the day time grow with the exact same slope, while in the night time the edge transmission times grow a bit slower. For the HTTP GET application instead, the trend is similar in both night time and day time (Figures~\ref{fig:night-app2-results} and~\ref{fig:day-app2-results}). The differences between the HTTP POST and the HTTP GET applications can be due to the different transmission technologies for downlink and uplink in LTE.

It has to be noted that the values of $\rho$ in all scenarios of Figure~\ref{fig:realdata-results} show a trend that differs from the one obtained when a UDP-based application was considered. For example, for each $RTT^C$ considered, Figure~\ref{fig:ratioincreasingpktsize_3D} shows that the values of $\rho$ increase as the amount of exchanged data increases. In the connection-oriented model this happens only for the HTTP POST application in the day time (Figure~\ref{fig:day-app1-results}), and still the trend is not the same as in Figure~\ref{fig:ratioincreasingpktsize_3D}. This can happen for mainly two reasons. Firstly, the connection-oriented experiments use the HTTP protocol implemented by existing tools that may introduce additional overhead on both the server and the Raspberry Pi.
Secondly, the TCP protocol provides several mechanisms to adapt the application throughput to the status of both the end-points and the network, while the UDP protocol lacks such mechanisms. As a consequence, the values of $T_{TX}$ and $T_{RX}$ computed for the connectionless model are independent from the server location, and the differences in the amount of energy consumed are dominated only by small differences in $T_W$ and $T_Q$. Instead, when a connection-oriented communication is considered, the throughput is affected by the RTT, which in turn depends on the server distance. 

\subsection{Evaluating the Impact of Multiple Clients}
\label{sec:multiple-clients}

\begin{figure}[!t]
    \centering
    \includegraphics[width=0.47\textwidth]{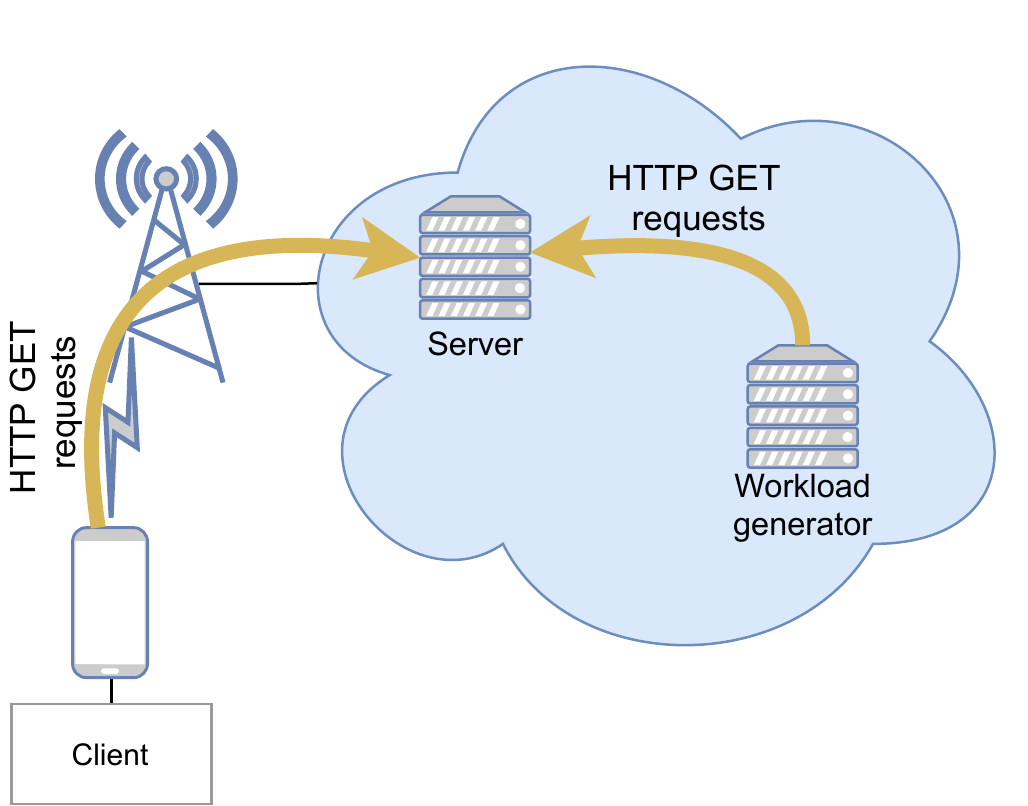}
    \caption{The setup used to collect $T_{TX}$, $T_W$, and $T_{RX}$ values when increasing the load on the server.}
    \label{fig:setup_concurrentconnections}
\end{figure}

In this section, we evaluate how the performance of an LTE device running a connection-oriented application can be affected by the presence of other devices interacting with the same server. To this purpose, we performed a set of experiments using the setup of Figure~\ref{fig:setup_concurrentconnections}. The specific goal was to collect the $T_{TX}$, $T_{RX}$, and $T_W$ values between the monitored client while the server, at the same time, communicates with other clients.
The monitored client is the one described in the previous sections. The server side is implemented on top of the \emph{Express} framework and it was hosted at the University of Pisa.
The client periodically sends HTTP GET requests toward the Express server, asking for a resource of fixed size (100 KB). An artificial workload was generated using \emph{autocannon}~\cite{autocannon}. The latter is a tool used to stress-test Web applications. Autocannon was executed on the same network where the Express server was located, to enable the execution of large numbers of requests. An increasing number of concurrent connections was used to simulate the presence of larger pools of clients. For each concurrent connection, autocannon executed GET requests as fast as possible.  
Finally, to exert a non-negligible computational load on the server the response was a randomly generated string.

\begin{algorithm}
    \caption{The procedure followed during the experiments.}
    \label{alg:concurrentconnections}
    \For{i $\in$ \{1, 2, 3, 4, 5, 6\}}{
        \For{c $\in$ \{0, 10, 50, 100, 150, 200\}}{
            Configure the cross-traffic generator\;
            
            \For{s $\in$ [$server_{EDGE}$, $server_{CLOUD}$]}{
                Start the tshark session\;
                Send an HTTP GET request to download a resource from server $s$\;
                Stop the tshark session and store the cap results\;
            }
        }
    }
\end{algorithm}

The experiments were performed according to the pseudo-code of Algorithm \ref{alg:concurrentconnections}.
At the beginning of each experiment, the workload generator is activated to send HTTP GET requests using $c$ concurrent connections. Then, the client starts a tshark session and sends an HTTP GET request to download a resource from the server. The collected trace is stored for later processing.  The last step is carried out for both the edge and the cloud configuration.
Finally, the workload generator is re-configured to use a higher amount of concurrent connections and the whole procedure is repeated. 

\begin{figure}[!t]
 \centering
    \includegraphics[width=0.47\textwidth]{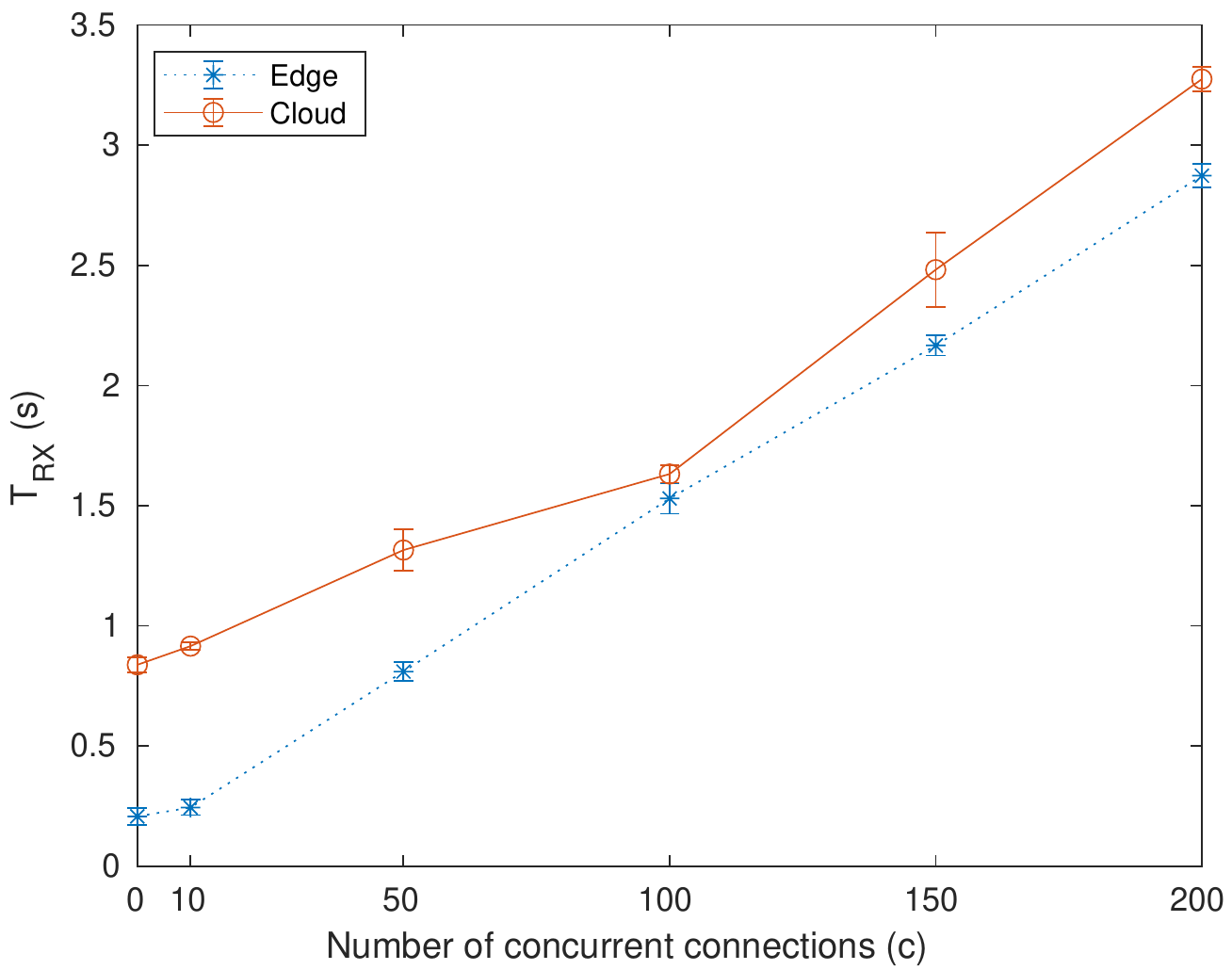}
    \caption{The mean time required to receive 100 KB considering an increasing server workload and a $\Delta$RTT of 200 milliseconds. The mean values have been computed upon 6 repetitions. The figure shows the 95\% confidence intervals. }
    \label{fig:realdatatempi-concurrentconnections}
\end{figure}

Figure \ref{fig:realdatatempi-concurrentconnections} shows the time needed to download 100 KB when varying the server workload. In particular, the curves in Figure~\ref{fig:realdatatempi-concurrentconnections} are the average values obtained with 6 repetitions.
As can be seen, the maximum difference between edge and cloud performance can be appreciated when there is no HTTP request sent by the cross-traffic generator. Conversely, edge and cloud performance are more similar when a high number of concurrent connections are considered. 
It should be noted that a high number of concurrent requests may lead to saturation of the links in the proximity of the server. Furthermore, since each response is generated randomly, a high number of requests inflict a high computational load on the server.
Hence, this result could indicate that communication latency tends to become less relevant when server resources are running low.

\begin{figure}[!t]
 \centering
    \includegraphics[width=0.47\textwidth]{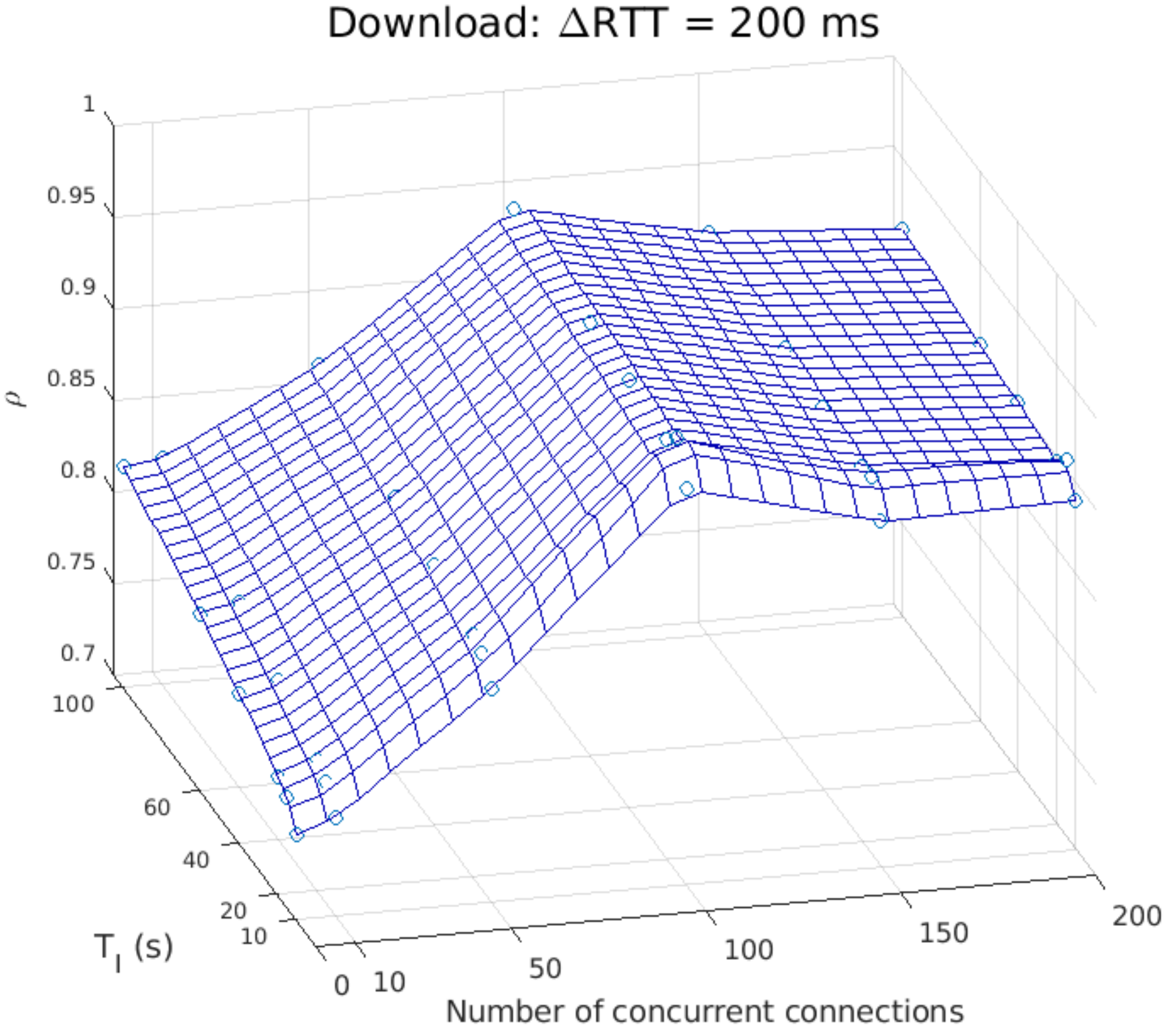}
    \caption{The ratio ($\rho$) between the $E_I^E$ and $E_I^C$, when varying the application period $T_I$ and the number of concurrent connections.}
    \label{fig:realdata-concurrentconnections-results}
\end{figure}

Finally, Figure~\ref{fig:realdata-concurrentconnections-results} show the ratio $\rho$ computed considering different values for the application period $T_I$ and different levels of concurrent connections. 
As can be seen, $\rho$ values are always lower than one. This indicates that the edge configuration is always the most convenient. Then, we can see that the $\rho$ values are coherent with the $T_{RX}$ values of Figure~\ref{fig:realdatatempi-concurrentconnections}. In fact, the lowest values of $\rho$ can be appreciated when $c$ is equal to zero. For such a value of $c$, the values of $T_ {RX}$ is significantly different for the two configurations.
Conversely, the highest $\rho$ values can be appreciated when $c$ is equal to 100, i.e. when the values of $T_ {RX}$ collected for the two configurations are close.
However, this result should not surprise since the interface is in CR during $T_{RX}$. Hence, differences in the time needed to receive the data have a relevant impact on the overall power consumption.

\section{Discussion}
\label{sec:discussion}

The results obtained show that, almost always, edge-based solutions allow saving energy on LTE terminal nodes. In particular, results highlight under what network conditions benefits are obtained compared to solutions that rely on a remote cloud.  Reducing the energy consumption on terminal devices is a fundamental step for obtaining greener communications and a more sustainable Internet. In addition, in some domains such as Industry 4.0 and Smart Cities, the number of devices belonging to a single organization can be in the order of hundreds or thousands. More efficient communication does not only bring reduced operational costs, but also easier management of battery operated devices. In some cases, the designer just has to push the application server to the edge to achieve some benefits, in other cases some tuning of application operational parameters can be useful (as shown in Section~\ref{sec:example}). 

Results have been obtained under a precise set of assumptions, in particular the request-response scheme, the periodic activity, and, in some cases, the use of HTTP as the application level protocol. While these assumptions can be frequently met for the considered scenario concerning smart devices, it is also possible that some applications could be characterized by a highly irregular communication pattern. These aspects have not been included in our model to keep reasonable the level of complexity and understand the main dynamics of the phenomenon. The obtained results pave the way for future work aimed at evaluating the impact of less regular communication patters -- such as when communication is event-diven-- and application-level protocols not based on HTTP.

\section{Related work}
\label{sec:relwork}

Our work, as mentioned, uses the energy model of the LTE interface defined in \cite{10.1145/2796314.2745875}, which in turn was based on \cite{10.1145/2307636.2307658}. 
In particular, in \cite{10.1145/2307636.2307658} the authors presented a method to infer the operational parameters of an LTE interface: by sending to the device packets with opportunely calibrated intervals and observing the time to respond, the transition times and the duty cycle of DRX modes were estimated. Both the model and the parameters were validated using an hardware power monitor~\cite{monsoon}, obtaining an error rate below $6\%$. In \cite{10.1145/2796314.2745875} the LTE power model was refined and used to estimate the energy expenditure of users in the wild. The goal of the study was to understand how the energy of a smartphone is depleted by its components: CPU, GPU, Wi-Fi, and the cellular interface. 
Data derived from many users showed that $24.4\%$ of the overall energy was spent on cellular communications. The study did not consider the energy benefits brought by edge computing when compared to centralized cloud, nor evaluated the impact of latency on energy consumption. However, the significant amount of energy spent by a smartphone for cellular communication proves the need for our work.

Other studies focused on finding the optimal DRX parameters, to balance the power-saving constraints with latency requirements. In \cite{7047910}, the problem was faced by using Markov chains, whereas in \cite{4657144} and \cite{s19030617} semi-Markov processes were used. Machine learning was used in \cite{9072614} to predict the optimal sleeping periods of an interface. All these works highlight the importance of reducing the energy needed by an LTE interface, but according to an orthogonal approach compared to our study. Our analysis highlights that the adoption of a network architecture based on edge computing is able to reduce the energy demands of client nodes. The benefits, in terms of energy, brought by edge computing were marginally evaluated in the past, and such benefits can be \emph{added} to the ones obtained by calibrating the LTE interface parameters. In other words, while the above literature studies how the parameters of operation of an LTE interface can be tuned, we evaluated the impact of external elements, in particular the placement of the server the mobile node communicates with. 

Some works focus on edge computing and energy consumption. The majority of studies at the intersection of the two topics are related to computation offloading, i.e. moving CPU-intensive tasks from mobile devices to the edge. In fact, smart devices are characterized by limited computational resources and delegating tasks to an edge server can be helpful in many domains, such as authentication \cite{8786231}, e-health \cite{RAY20191}, and workflow applications \cite{10.1186/s13638-019-1526-x}.
In addition, smart devices are typically battery-powered, thus, offloading part of the computation may bring benefits in terms of required energy. 
In \cite{9044818} a technique based on cooperative task offloading and caching in a MEC environment is presented. The technique minimizes the energy consumption while guaranteeing proper security levels. 
The performance of a three-tier fog network for IoT applications has been studied in \cite{7286781}, considering power consumption, service latency, and CO$_2$ emission metrics. The overall power consumption has been split into the power consumed to forward packets, perform computation, store data, and migrate applications in the cloud. 
Task offloading mechanisms for hierarchical edge computing networks have been analyzed in \cite{10.1155/2020/6098786}, and \cite{8234573} where the offloading strategy is optimized to take into account both the energy costs and the task execution latency. In hierarchical edge computing networks there are different layers of edge servers, more or less powerful in terms of computational power, and more or less distant from the end users. Other works on the optimal energy-aware offloading schemes in a MEC environment can been found in \cite{MAZOUZI2019132} and \cite{LI2021183}.
However, all the above studies are based on the idea of moving part of the computation onto edge nodes. We, on the contrary, do not assume any transfer of tasks from terminal nodes to edge servers. Our study makes clear which benefits can be obtained when communicating with servers at different distances in the latency space. This information could be useful to include a refined model of communication costs, in terms of energy, in many of the studies concerning task offloading. The reader is also forwarded to \cite{JIANG2020556} for a survey about energy aware edge computing, where approaches at different levels are summarized, from specific hardware designs to middleware. 

An experimental evaluation of the benefits of offloading computation to edge- and cloud-based servers is presented in \cite{10.1145/2967360.2967369}. Different applications characterized by different levels of computing requirements were tested. The results show that, in many cases, an edge server can reduce both energy consumption and response time compared to keeping the computation local to the device or using a more distant server in the cloud. Also such work focuses on computation offloading. Our work on the contrary does not assume any offloading of computation. Not only our results are related to pure communication, but also provide an indication of \emph{why} the edge (or the cloud) provides benefits in terms of needed energy. For the results provided in Section~\ref{sec:udpresults} and~\ref{sec:tcpresults}, we highlighted how energy is spent during the different phases of communication. This, in turn, provides a conceptual framework that allows the application designers to make their applications more energy efficient.

\section{Conclusions}
\label{sec:conclusions}
The introduction of edge nodes close to the terminal devices provides several advantages such as reduced latency, higher bandwidth, reduced traffic in the core network, and distributed workload among edge and cloud servers. Further, deploying application components on edge nodes may diminish the energy consumption of terminal devices, which generally are battery-powered. 

We assessed the energy consumption of a terminal node operating in an edge computing infrastructure. Starting from a FSM-based model of the LTE interface, we computed the energy consumption on the terminal nodes, considering a simple client-server application. We computed energy consumption in a connectionless communication scenario analytically using the FSM, and the energy consumption in a connection-oriented communication scenario by introducing real-world measurements into the model.
The results show that the edge server is always convenient in a connection-oriented scenario. This result can be mainly ascribed to the dependency of the TCP throughput from the RTT between the client and the server.
When the client interacts with a remote cloud server, the throughput grows slower as it is influenced by the higher RTT experienced between the two end-points. As a result, a longer amount of time is required to send (or receive) the data. This forces the client interface to stay longer in the CR state, which is characterized by high power consumption. 
On the other hand, the power consumption in a connectionless communication scenario is less affected by the location of the server. The edge-based solution is still generally more favorable, but in some cases the cloud-based one proved to be a better choice. As connectionless communication protocols, such as UDP, lack of any flow and congestion control mechanisms, the time required to send (or receive) some data is independent of the server location. Differences can be found only in idle periods, where no data is transmitted, and the interface may be in a state different than CR with lower energy consumption.

\section*{Acknowledgment}

This work is partially funded by the Italian
Ministry of Education and Research (MIUR) in the framework of the CrossLab project (Departments of Excellence). The views expressed are solely those of the authors.

\balance
\bibliographystyle{elsarticle-num}
\bibliography{bibliography}

\begin{thebibliography}{10}
\expandafter\ifx\csname url\endcsname\relax
  \def\url#1{\texttt{#1}}\fi
\expandafter\ifx\csname urlprefix\endcsname\relax\def\urlprefix{URL }\fi
\expandafter\ifx\csname href\endcsname\relax
  \def\href#1#2{#2} \def\path#1{#1}\fi

\bibitem{8685768}
X.~Zhang, H.~Chen, Y.~Zhao, Z.~Ma, Y.~Xu, H.~Huang, H.~Yin, D.~O. Wu,
  {Improving Cloud Gaming Experience through Mobile Edge Computing}, IEEE
  Wireless Communications 26~(4) (2019) 178--183.

\bibitem{7980118}
T.~Braud, F.~H. Bijarbooneh, D.~Chatzopoulos, P.~Hui, {Future Networking
  Challenges: The Case of Mobile Augmented Reality}, in: 2017 IEEE 37th
  International Conference on Distributed Computing Systems (ICDCS), 2017, pp.
  1796--1807.

\bibitem{8515152}
F.~Giust, V.~Sciancalepore, D.~Sabella, M.~C. Filippou, S.~Mangiante,
  W.~Featherstone, D.~Munaretto, {Multi-Access Edge Computing: The Driver
  Behind the Wheel of 5G-Connected Cars}, IEEE Communications Standards
  Magazine 2~(3) (2018) 66--73.

\bibitem{8567664}
J.~Wang, Z.~Feng, Z.~Chen, S.~George, M.~Bala, P.~Pillai, S.-W. Yang,
  M.~Satyanarayanan, {Bandwidth-Efficient Live Video Analytics for Drones Via
  Edge Computing}, in: 2018 IEEE/ACM Symposium on Edge Computing (SEC), 2018,
  pp. 159--173.

\bibitem{giust2018mec}
F.~Giust, G.~Verin, K.~Antevski, J.~Chou, Y.~Fang, W.~Featherstone, F.~Fontes,
  D.~Frydman, A.~Li, A.~Manzalini, et~al., {MEC deployments in 4G and evolution
  towards 5G}, ETSI White paper 24~(2018) (2018) 1--24.

\bibitem{8969038}
R.~Srinivasa, N.~K.~S. Naidu, S.~Maheshwari, C.~Bharathi, A.~R. Hemanth~Kumar,
  {Minimizing Latency for 5G Multimedia and V2X Applications using Mobile Edge
  Computing}, in: 2019 2nd International Conference on Intelligent
  Communication and Computational Techniques (ICCT), 2019, pp. 213--217.

\bibitem{7740617}
G.~B. Fioccola, R.~Sommese, I.~Tufano, R.~Canonico, G.~Ventre, Polluino: An
  efficient cloud-based management of iot devices for air quality monitoring,
  in: 2016 IEEE 2nd International Forum on Research and Technologies for
  Society and Industry Leveraging a better tomorrow (RTSI), 2016, pp. 1--6.
\newblock \href {https://doi.org/10.1109/RTSI.2016.7740617}
  {\path{doi:10.1109/RTSI.2016.7740617}}.

\bibitem{MASUD2020215}
M.~Masud, G.~Muhammad, H.~Alhumyani, S.~S. Alshamrani, O.~Cheikhrouhou,
  S.~Ibrahim, M.~S. Hossain,
  \href{https://www.sciencedirect.com/science/article/pii/S0140366419312988}{Deep
  learning-based intelligent face recognition in iot-cloud environment},
  Computer Communications 152 (2020) 215--222.
\newblock \href {https://doi.org/https://doi.org/10.1016/j.comcom.2020.01.050}
  {\path{doi:https://doi.org/10.1016/j.comcom.2020.01.050}}.
\newline\urlprefix\url{https://www.sciencedirect.com/science/article/pii/S0140366419312988}

\bibitem{yang2016iot}
Z.~Yang, Q.~Zhou, L.~Lei, K.~Zheng, W.~Xiang, An iot-cloud based wearable ecg
  monitoring system for smart healthcare, Journal of medical systems 40~(12)
  (2016) 1--11.

\bibitem{RAY201635}
P.~P. Ray,
  \href{https://www.sciencedirect.com/science/article/pii/S2314728816300149}{A
  survey of iot cloud platforms}, Future Computing and Informatics Journal
  1~(1) (2016) 35--46.
\newblock \href {https://doi.org/https://doi.org/10.1016/j.fcij.2017.02.001}
  {\path{doi:https://doi.org/10.1016/j.fcij.2017.02.001}}.
\newline\urlprefix\url{https://www.sciencedirect.com/science/article/pii/S2314728816300149}

\bibitem{7116451}
H.-L. Truong, S.~Dustdar, Principles for engineering iot cloud systems, IEEE
  Cloud Computing 2~(2) (2015) 68--76.
\newblock \href {https://doi.org/10.1109/MCC.2015.23}
  {\path{doi:10.1109/MCC.2015.23}}.

\bibitem{botta2016integration}
A.~Botta, W.~De~Donato, V.~Persico, A.~Pescap{\'e}, Integration of cloud
  computing and internet of things: a survey, Future generation computer
  systems 56 (2016) 684--700.

\bibitem{samie2016computation}
F.~Samie, V.~Tsoutsouras, L.~Bauer, S.~Xydis, D.~Soudris, J.~Henkel,
  Computation offloading and resource allocation for low-power iot edge
  devices, in: 2016 IEEE 3rd World Forum on Internet of Things (WF-IoT), IEEE,
  2016, pp. 7--12.

\bibitem{morabito2018consolidate}
R.~Morabito, V.~Cozzolino, A.~Y. Ding, N.~Beijar, J.~Ott, Consolidate iot edge
  computing with lightweight virtualization, Ieee network 32~(1) (2018)
  102--111.

\bibitem{7469991}
W.~Shi, S.~Dustdar, The promise of edge computing, Computer 49~(5) (2016)
  78--81.
\newblock \href {https://doi.org/10.1109/MC.2016.145}
  {\path{doi:10.1109/MC.2016.145}}.

\bibitem{8466364}
B.~Chen, J.~Wan, A.~Celesti, D.~Li, H.~Abbas, Q.~Zhang, Edge computing in
  iot-based manufacturing, IEEE Communications Magazine 56~(9) (2018) 103--109.
\newblock \href {https://doi.org/10.1109/MCOM.2018.1701231}
  {\path{doi:10.1109/MCOM.2018.1701231}}.

\bibitem{10.1145/2967360.2967369}
W.~Hu, Y.~Gao, K.~Ha, J.~Wang, B.~Amos, Z.~Chen, P.~Pillai, M.~Satyanarayanan,
  {Quantifying the Impact of Edge Computing on Mobile Applications}, in:
  Proceedings of the 7th ACM SIGOPS Asia-Pacific Workshop on Systems, APSys
  '16, Association for Computing Machinery, New York, NY, USA, 2016.

\bibitem{10.1145/3132211.3134446}
B.~Qi, L.~Kang, S.~Banerjee, \href{https://doi.org/10.1145/3132211.3134446}{A
  vehicle-based edge computing platform for transit and human mobility
  analytics}, in: Proceedings of the Second ACM/IEEE Symposium on Edge
  Computing, SEC '17, Association for Computing Machinery, New York, NY, USA,
  2017.
\newblock \href {https://doi.org/10.1145/3132211.3134446}
  {\path{doi:10.1145/3132211.3134446}}.
\newline\urlprefix\url{https://doi.org/10.1145/3132211.3134446}

\bibitem{Savaglio19:iot}
C.~Savaglio, G.~Campisano, G.~di~Fatta, G.~Fortino, {IoT Services Deployment
  over Edge vs Cloud Systems: a Simulation-based Analysis}, in: IEEE INFOCOM
  2019 - IEEE Conference on Computer Communications Workshops (INFOCOM WKSHPS),
  2019, pp. 554--559.
\newblock \href {https://doi.org/10.1109/INFCOMW.2019.8845305}
  {\path{doi:10.1109/INFCOMW.2019.8845305}}.

\bibitem{Heredia19:edge}
A.~Heredia, G.~Barros-Gavilanes,
  \href{https://doi.org/10.1117/12.2533380}{{Edge vs. cloud computing: where to
  do image processing for surveillance?}}, in: J.~Dijk (Ed.), Artificial
  Intelligence and Machine Learning in Defense Applications, Vol. 11169,
  International Society for Optics and Photonics, SPIE, 2019, pp. 20 -- 32.
\newblock \href {https://doi.org/10.1117/12.2533380}
  {\path{doi:10.1117/12.2533380}}.
\newline\urlprefix\url{https://doi.org/10.1117/12.2533380}

\bibitem{Silva19:investigating}
P.~Silva, A.~Costan, G.~Antoniu, {Investigating Edge vs. Cloud Computing
  Trade-offs for Stream Processing}, in: 2019 IEEE International Conference on
  Big Data (Big Data), 2019, pp. 469--474.
\newblock \href {https://doi.org/10.1109/BigData47090.2019.9006139}
  {\path{doi:10.1109/BigData47090.2019.9006139}}.

\bibitem{Nikolaou19:evaluation}
P.~Nikolaou, Y.~Sazeides, A.~Lampropoulos, D.~Guilhot, A.~Bartoli,
  G.~Papadimitriou, A.~Chatzidimitriou, D.~Gizopoulos, K.~Tovletoglou,
  L.~Mukhanov, G.~Karakonstantis, {On the Evaluation of the
  Total-Cost-of-Ownership Trade-offs in Edge vs Cloud deployments: A
  Wireless-Denial-of-Service Case Study}, IEEE Transactions on Sustainable
  Computing (2019) 1--1\href {https://doi.org/10.1109/TSUSC.2019.2894018}
  {\path{doi:10.1109/TSUSC.2019.2894018}}.

\bibitem{Luckow21:exploring}
A.~Luckow, K.~Rattan, S.~Jha, {Exploring Task Placement for Edge-to-Cloud
  Applications using Emulation}, in: 2021 IEEE 5th International Conference on
  Fog and Edge Computing (ICFEC), 2021, pp. 79--83.
\newblock \href {https://doi.org/10.1109/ICFEC51620.2021.00019}
  {\path{doi:10.1109/ICFEC51620.2021.00019}}.

\bibitem{6629847}
N.~Nikaein, M.~Laner, K.~Zhou, P.~Svoboda, D.~Drajic, M.~Popovic, S.~Krco,
  Simple traffic modeling framework for machine type communication, in: ISWCS
  2013; The Tenth International Symposium on Wireless Communication Systems,
  2013, pp. 1--5.

\bibitem{10.1145/2796314.2745875}
X.~Chen, N.~Ding, A.~Jindal, Y.~C. Hu, M.~Gupta, R.~Vannithamby, {Smartphone
  Energy Drain in the Wild: Analysis and Implications}, SIGMETRICS Perform.
  Eval. Rev. 43~(1) (2015) 151–164.

\bibitem{10.1145/2307636.2307658}
J.~Huang, F.~Qian, A.~Gerber, Z.~M. Mao, S.~Sen, O.~Spatscheck, {A Close
  Examination of Performance and Power Characteristics of 4G LTE Networks}, in:
  Proceedings of the 10th International Conference on Mobile Systems,
  Applications, and Services, MobiSys '12, Association for Computing Machinery,
  New York, NY, USA, 2012, p. 225–238.

\bibitem{CAIAZZA2021108140}
C.~Caiazza, C.~Cicconetti, V.~Luconi, A.~Vecchio, {Measurement-driven design
  and runtime optimization in edge computing: Methodology and tools}, Computer
  Networks 194 (2021) 108140.

\bibitem{10.1145/263932.264023}
M.~Mathis, J.~Semke, J.~Mahdavi, T.~Ott, {The Macroscopic Behavior of the TCP
  Congestion Avoidance Algorithm}, SIGCOMM Comput. Commun. Rev. 27~(3) (1997)
  67–82.

\bibitem{10.1145/285243.285291}
J.~Padhye, V.~Firoiu, D.~Towsley, J.~Kurose, {Modeling TCP Throughput: A Simple
  Model and Its Empirical Validation}, SIGCOMM Comput. Commun. Rev. 28~(4)
  (1998) 303–314.

\bibitem{cardwell1998modeling}
N.~Cardwell, S.~Savage, T.~Anderson, {Modeling the performance of short TCP
  connections}, Techical Report (1998).

\bibitem{Bao2010:model}
W.~Bao, V.~W.~S. Wong, V.~C.~M. Leung, {A Model for Steady State Throughput of
  TCP CUBIC}, in: 2010 IEEE Global Telecommunications Conference GLOBECOM 2010,
  2010, pp. 1--6.

\bibitem{autocannon}
autocannon:an http/1.1 benchmarking tool,
  \url{https://github.com/mcollina/autocannon}, accessed: 2022-02-18.

\bibitem{monsoon}
Monsoon power monitor, \url{https://www.msoon.com/high-voltage-power-monitor},
  accessed: 2022-02-18.

\bibitem{7047910}
C.~{Tseng}, H.~{Wang}, F.~{Kuo}, K.~{Ting}, H.~{Chen}, G.~{Chen}, {Delay and
  Power Consumption in LTE/LTE-A DRX Mechanism With Mixed Short and Long
  Cycles}, IEEE Transactions on Vehicular Technology 65~(3) (2016) 1721--1734.

\bibitem{4657144}
L.~{Zhou}, H.~{Xu}, H.~{Tian}, Y.~{Gao}, L.~{Du}, L.~{Chen}, {Performance
  Analysis of Power Saving Mechanism with Adjustable DRX Cycles in 3GPP LTE},
  in: 2008 IEEE 68th Vehicular Technology Conference, 2008, pp. 1--5.

\bibitem{s19030617}
Y.~Mehmood, L.~Zhang, A.~Förster, {Power Consumption Modeling of Discontinuous
  Reception for Cellular Machine Type Communications}, Sensors 19~(3) (2019).

\bibitem{9072614}
P.~{Brand}, J.~{Falk}, J.~{Ah Sue}, J.~{Brendel}, R.~{Hasholzner}, J.~{Teich},
  {Adaptive Predictive Power Management for Mobile LTE Devices}, IEEE
  Transactions on Mobile Computing (2020) 1--1.

\bibitem{8786231}
P.~Hao, X.~Wang, {Integrating PHY Security Into NDN-IoT Networks By Exploiting
  MEC: Authentication Efficiency, Robustness, and Accuracy Enhancement}, IEEE
  Transactions on Signal and Information Processing over Networks 5~(4) (2019)
  792--806.

\bibitem{RAY20191}
P.~P. Ray, D.~Dash, D.~De,
  \href{https://www.sciencedirect.com/science/article/pii/S1084804519301651}{Edge
  computing for internet of things: A survey, e-healthcare case study and
  future direction}, Journal of Network and Computer Applications 140 (2019)
  1--22.
\newblock \href {https://doi.org/https://doi.org/10.1016/j.jnca.2019.05.005}
  {\path{doi:https://doi.org/10.1016/j.jnca.2019.05.005}}.
\newline\urlprefix\url{https://www.sciencedirect.com/science/article/pii/S1084804519301651}

\bibitem{10.1186/s13638-019-1526-x}
K.~Peng, M.~Zhu, Y.~Zhang, L.~Liu, J.~Zhang, V.~C.~M. Leung, L.~Zheng,
  \href{https://doi.org/10.1186/s13638-019-1526-x}{An energy- and cost-aware
  computation offloading method for workflow applications in mobile edge
  computing}, EURASIP Journal on Wireless Communications and Networking
  2019~(1) (2019) 207.
\newblock \href {https://doi.org/10.1186/s13638-019-1526-x}
  {\path{doi:10.1186/s13638-019-1526-x}}.
\newline\urlprefix\url{https://doi.org/10.1186/s13638-019-1526-x}

\bibitem{9044818}
M.~I.~A. Zahed, I.~Ahmad, D.~Habibi, Q.~V. Phung, {Green and Secure Computation
  Offloading for Cache-Enabled IoT Networks}, IEEE Access 8 (2020)
  63840--63855.

\bibitem{7286781}
S.~Sarkar, S.~Chatterjee, S.~Misra, {Assessment of the Suitability of Fog
  Computing in the Context of Internet of Things}, IEEE Transactions on Cloud
  Computing 6~(01) (2018) 46--59.

\bibitem{10.1155/2020/6098786}
Y.~Pei, Z.~Peng, Z.~Wang, H.~Wang, {Energy-Efficient Mobile Edge Computing:
  Three-Tier Computing under Heterogeneous Networks}, Wirel. Commun. Mob.
  Comput. 2020 (Jan. 2020).

\bibitem{8234573}
J.~{Zhang}, X.~{Hu}, Z.~{Ning}, E.~C.~. {Ngai}, L.~{Zhou}, J.~{Wei},
  J.~{Cheng}, B.~{Hu}, {Energy-Latency Tradeoff for Energy-Aware Offloading in
  Mobile Edge Computing Networks}, IEEE Internet of Things Journal 5~(4) (2018)
  2633--2645.

\bibitem{MAZOUZI2019132}
H.~Mazouzi, K.~Boussetta, N.~Achir,
  \href{https://www.sciencedirect.com/science/article/pii/S0140366419301720}{Maximizing
  mobiles energy saving through tasks optimal offloading placement in two-tier
  cloud: A theoretical and an experimental study}, Computer Communications 144
  (2019) 132--148.
\newblock \href {https://doi.org/https://doi.org/10.1016/j.comcom.2019.05.017}
  {\path{doi:https://doi.org/10.1016/j.comcom.2019.05.017}}.
\newline\urlprefix\url{https://www.sciencedirect.com/science/article/pii/S0140366419301720}

\bibitem{LI2021183}
X.~Li, L.~Zhao, K.~Yu, M.~Aloqaily, Y.~Jararweh,
  \href{https://www.sciencedirect.com/science/article/pii/S0140366421001419}{A
  cooperative resource allocation model for iot applications in mobile edge
  computing}, Computer Communications 173 (2021) 183--191.
\newblock \href {https://doi.org/https://doi.org/10.1016/j.comcom.2021.04.005}
  {\path{doi:https://doi.org/10.1016/j.comcom.2021.04.005}}.
\newline\urlprefix\url{https://www.sciencedirect.com/science/article/pii/S0140366421001419}

\bibitem{JIANG2020556}
C.~Jiang, T.~Fan, H.~Gao, W.~Shi, L.~Liu, C.~Cérin, J.~Wan,
  \href{https://www.sciencedirect.com/science/article/pii/S014036641930831X}{Energy
  aware edge computing: A survey}, Computer Communications 151 (2020) 556--580.
\newblock \href {https://doi.org/https://doi.org/10.1016/j.comcom.2020.01.004}
  {\path{doi:https://doi.org/10.1016/j.comcom.2020.01.004}}.
\newline\urlprefix\url{https://www.sciencedirect.com/science/article/pii/S014036641930831X}

\end{thebibliography}

\clearpage

\end{document}